\documentclass[twocolumn]{aa}
\usepackage{graphicx}
\usepackage{amsbsy}
\usepackage{natbib}
\usepackage[usenames]{color}
\usepackage{txfonts}
\newcommand{\de}{\mathrm{d}}
\newcommand{\dpa}{\partial}

\DeclareMathSymbol{\varOmega}{\mathord}{letters}{"0A}
\DeclareMathSymbol{\varSigma}{\mathord}{letters}{"06}
\DeclareMathSymbol{\varPsi}{\mathord}{letters}{"09}

\newcommand{\Eq}[1]{Eq.~(\ref{#1})}
\newcommand{\Eqs}[2]{Eqs.~(\ref{#1}) and~(\ref{#2})}
\newcommand{\Eqss}[2]{Eqs.~(\ref{#1})-(\ref{#2})}
\newcommand{\App}[1]{Appendix~\ref{#1}}

\newcommand{\Fig}[1]{Fig.~\ref{#1}}
\newcommand{\Figs}[2]{Figs.~\ref{#1} and \ref{#2}}
\newcommand{\Tab}[1]{Table \ref{#1}}
\begin{document}
\title{A coagulation--fragmentation model\\for the turbulent growth and
destruction of preplanetesimals}
\titlerunning{The growth and destruction of preplanetesimals}

\author{Anders Johansen, Frithjof Brauer, Cornelis Dullemond, Hubert Klahr 
\and Thomas Henning\inst{1}}
\authorrunning{Johansen et al.}

\offprints{A. Johansen}

\institute{Max-Planck-Institut f\"ur Astronomie,
  K\"onigstuhl 17, 69117 Heidelberg, Germany\\
  \email{johansen@mpia.de}}

\date{\today}

\abstract{
To treat the problem of growing protoplanetary disc solids across the meter
barrier, we consider a very simplified two-component coagulation-fragmentation
model that consists of macroscopic boulders and smaller dust grains, the latter
being the result of catastrophic collisions between the boulders. Boulders in
turn increase their radii by sweeping up the dust fragments. An analytical
solution to the dynamical equations predicts that growth by
coagulation-fragmentation can be efficient and allow agglomeration of
10-meter-sized objects within the time-scale of the radial drift. These results
are supported by computer simulations of the motion of boulders and fragments
in 3-D time-dependent magnetorotational turbulence. Allowing however the
fragments to diffuse freely out of the sedimentary layer of boulders reduces
the density of both boulders and fragments in the mid-plane, and thus also the
growth of the boulder radius, drastically. The reason is that the turbulent
diffusion time-scale is so much shorter than the collisional time-scale that
dust fragments leak out of the mid-plane layer before they can be swept up by
the boulders there. Our conclusion that coagulation-fragmentation is not an
efficient way to grow across the meter barrier in fully turbulent
protoplanetary discs confirms recent results by Brauer, Dullemond, \& Henning
who solved the coagulation equation in a parameterised turbulence model with
collisional fragmentation, cratering, radial drift, and a range of particle
sizes. We find that a relatively small population of boulders in a sedimentary
mid-plane layer can populate the entire vertical extent of the disc with small
grains and that these grains are not first generation dust, but have been
through several agglomeration-destruction cycles during the simulations.
}

\maketitle

\section{Introduction}

The formation of km-sized planetesimals from $\mu$m-sized dust grains is a
long-standing challenge of planet formation. The problem is complicated by the
interplay of an array of different physics -- most notably the turbulence of
protoplanetary discs and the sticking and collisional destruction of solids of
different sizes
\citep{Chokshi+etal1993,WeidenschillingCuzzi1993,DominikTielens1997,BlumWurm2000,Henning+etal2006}.

Young stars are known to receive mass from the inner part of their
circumstellar disc \citep{Bertout+etal1988}. The cause of such accretion is
most likely that protoplanetary discs are turbulent. The source of turbulence
must in this connection be a Keplerian shear instability \citep[such as the
magnetorotational instability, see][]{BalbusHawley1998} or self-gravity if the
disc is massive enough \citep{BalbusPapaloizou1999,Gammie2001,LodatoRice2004}.
There are nevertheless significant problems with both these sources of
accretion: the ionisation fraction of protoplanetary discs at 1--20 AU from the
star may be too low for the magnetorotational instability to operate
\citep{Gammie1996,Semenov+etal2004}, while only a minor fraction of discs are
expected to be massive enough to be gravitationally unstable
\citep{Beckwith+etal1990}. Thus turbulence is often treated as a free parameter
in protoplanetary disc models, parameterised through an $\alpha$-value (i.e.\
turbulent viscosity) anywhere from $\alpha=10^{-6}$ up to as high as
$\alpha=0.1$.

Collisions between $\mu$m-sized dust monomers leads to the formation of dust
aggregates under a range of conditions \citep{BlumWurm2000}. But larger bodies
have poor sticking properties and a lower threshold for collisional destruction
\citep{Chokshi+etal1993,Benz2000}. The sticking problem is especially acute for
m-sized boulders. The strength of these bodies is very low, while collision
speeds, induced by the turbulent gas and by the differential settling and
radial drift, are high
\citep{Voelk+etal1980,Mizuno+etal1988,Weidenschilling1997}. The slightly
sub-Keplerian gas acts as a constant head wind on the boulders, draining them
of angular momentum and causing them to drift radially through the disc
\citep{Weidenschilling1977}. The drift rate is approximately proportional to
the radius of the boulders, introducing a differential radial drift that can be
as strong as 50 m/s in difference between bodies of 1 m and 10 cm in size.

Modelling the growth of solids in protoplanetary discs requires solution of the
coagulation equation (or Smoluchowski equation) that governs the time evolution
of a size distribution of solids. Pioneering work on the numerical solution of
the coagulation equation in a planetesimal formation context was done by
\cite{Weidenschilling1984} who found that particle growth in turbulent discs is
efficient because the relative speeds induced by the turbulence give high
collision rates. Collisional fragmentation nevertheless halts growth when the
solids reach sizes of a few cm (pebbles). These early models were improved in
\cite{Weidenschilling1997} to include many more bins in the vertical direction
and in the particle radius. Considering the formation of comets at 30 AU from
the proto-Sun, Weidenschilling concluded that coagulation could in principle
explain the growth all the way to km-sized planetesimals, although for disc
models where the turbulence is purely induced by the sedimentation, so that
very high particle densities occur in the mid-plane. The sticking efficiency
was also assumed to be high, in some contrast with later models of boulder
collisions \citep{Benz2000,Schaefer+etal2007}, and collisional fragmentation
was kept at a minimum by assuming a constant specific kinetic energy threshold
for fragmentation. In this approach bodies of different sizes may collide at
very high speeds without destroying each other (see Appendix F of
\citealt{Weidenschilling1997}).

\cite{DullemondDominik2005} presented simulations similar to
\cite{Weidenschilling1997} and found numerically that the size distribution of
solids can get into a balance between collisional fragmentation and
coagulation. The first simulations to include the full radial extent of a
protoplanetary disc were presented recently by Brauer, Dullemond, \& Henning
(2008, hereafter BDH)\nocite{Brauer+etal2008}. BDH found that the meter barrier
is a genuine problem for planetesimal formation, both because macroscopic
bodies destroy each other in catastrophic collisions and because radial drift
sends macroscopic bodies into the inner nebula where they are lost from the
planet formation process. Radial drift is not only a problem for m-sized
boulders: over the life-times of protoplanetary discs \citep[millions of years,
see e.g.][]{Bouwman+etal2006} even mm- and cm-sized pebbles have significant
drift and are emptied from the outer parts of the nebula
\citep{TakeuchiLin2002,Brauer+etal2007}.

In this paper we isolate the effect of collisional fragmentation and subsequent
sweep-up of fragments on the growth of boulders. We do not solve the full
coagulation equation as in BDH, but simplify the size distribution to
effectively two bins -- small dust fragments and large boulders -- in order to
make the particle growth tractable in a 3-D simulation with magnetised
time-dependent turbulence.

The paper is structured as follows. In \S\ref{s:analytics} we describe our
simplified two-component model of boulders and fragments in detail and find an
equilibrium solution to the dynamical equations. In \S\ref{s:simulations} we
describe the numerical simulations that will be used to evolve the dynamical
equations. A local corotating box is considered, and turbulence is produced by
the magnetorotational instability. Dust fragments are treated as a passive
scalar, while the boulders are treated as individual superparticles with two
internal degrees of freedom: the number density of actual particles inside each
superparticle and the average radius of the constituent boulders. In the next
section, \S\ref{s:nosed}, we present simulations where the dust fragments are
not allowed to leave the boulder layer. The system quickly evolves towards the
equilibrium state found in \S\ref{s:analytics} with rapid growth of the
boulders of a few mm per orbit. In \S\ref{s:diffusion} we briefly turn to the
analytical model again and see how the diffusion of dust fragments from the
boulder layer affects the equilibrium solution. The turbulent gas transports
fragments out of the mid-plane, and thus the growth rate of the boulders is
reduced. This is confirmed in computer simulations where boulders lie in a thin
layer around the mid-plane, presented in \S\ref{s:sed}. The dust fragments
spread quickly out of the mid-plane and the growth rate of the boulders is
reduced by a factor of 10, so that the growth can no longer compete with the
radial drift. The whole disc is filled with dust fragments that are formed in
the thin mid-plane layer. We conclude on our results in \S\ref{s:conclusions}
and speculate about ways to get coagulation-fragmentation growth back on
track. We find that radial drift, not collisional fragmentation, is the more
serious problem for coagulation-fragmentation growth and discuss processes to
reduce or stop radial drift in actual protoplanetary discs.

\section{Coagulation-fragmentation model}
\label{s:analytics}

We consider a simple two-species model of solids in a protoplanetary disc.
Species 1 consists of tiny dust grains with mass $m_1$ and number density
$n_1$, while species 2 consists of macroscopic boulders with mass $m_2$ and
number density $n_2$. Here ``dust grains'' are defined as being so small that
they couple to the gas on a time-scale that is much shorter than an orbital
time, while ``boulders'' are solid bodies with sizes somewhere between 10 cm
and 10 m.

We assume for simplicity that
\begin{enumerate}
  \item Collisions between the tiny grains are insignificant compared to the
    sweep-up of the grains by the boulders.
  \item Collisions between a boulder and a dust grain always lead to the
    incorporation of the grain into the boulder.
  \item Collisions between the boulders lead to a complete destruction of the
    colliding bodies. The entire mass then ends up in tiny grains.
\end{enumerate}
Thus we do not treat the problem of how to form boulders in the first
place, but focus on how they grow by sweeping up dust. We also ignore effects
like cratering in our treatment of boulder-dust collisions. Although the actual
growth of boulders across the metre-barrier will likely involve a combination of
many different aspects of collision physics (and also self-gravity), we will in
this paper instead aim at gaining insight into the pure problem of collisional
fragmentation of equal-sized boulders and sweep-up of small dust grains in a
turbulent environment. Assuming an impact strength of zero for the boulders and
perfect sticking between boulders and dust grains may not be entirely
realistic, but this allows us to simplify our model greatly.

We can write up the dynamical equations for the number densities of dust and
boulders and for the mass of the individual boulders,
\begin{equation}\label{eq:dn1dt}
  \frac{\dpa n_1}{\dpa t} = -n_1 n_2 \sigma_{12} v_{12} +
      n_2^2 \sigma_{22} v_{22} \frac{m_2}{m_1} \, ,
\end{equation}
% Mental note: v looks like nu here.
\begin{equation}\label{eq:dn2dt}
  \frac{\dpa n_2}{\dpa t} = -n_2^2 \sigma_{22} v_{22} \, ,
\end{equation}
\begin{equation}\label{eq:dm2dt}
  \frac{\dpa m_2}{\dpa t} = \sigma_{12} v_{12} \rho_1 \, .
\end{equation}
Here $\sigma_{12}$ and $\sigma_{22}$ are the collisional cross sections for
boulder-grain and boulder-boulder collisions, respectively, while $v_{12}$ and
$v_{22}$ are the corresponding collision speeds. We furthermore introduced the
bulk density of dust fragments $\rho_1=n_1 m_1$. We assume next that the
grains and the boulders are spheres with radius $a_1$ and $a_2$, respectively,
and that $a_1 \ll a_2$. The dynamical equation for $m_2$ can then be turned
into a dynamical equation for the radius $a_2$,
\begin{equation}\label{eq:da2dt}
  \frac{\dpa a_2}{\dpa t} = \frac{\rho_1}{4 \rho_\bullet} v_{12} \, ,
\end{equation}
with $\rho_\bullet$ referring to the material density of the solids.

\subsection{Equilibrium limit}

\Eq{eq:dn1dt} has the equilibrium solution
\begin{equation}\label{eq:rhorat_equi}
  \frac{\rho_1}{\rho_2} = 4 \frac{v_{22}}{v_{12}} \, ,
\end{equation}
where we define the bulk densities\footnote{We use the term ``bulk density''
throughout this paper to refer to the total mass of solid material in a given
volume divided by the volume, i.e.\ including the void between the solids.}
$\rho_1=n_1 m_1$, $\rho_2=n_2 m_2$ and set $\sigma_{22}\approx4\sigma_{12}$
under the assumption that the contribution of the small grains to the cross
section $\sigma_{12}$ is vanishing. We have also assumed, by setting the
collisional radius of a boulder to twice its physical radius, that all impact
parameters lead to destruction, even if the boulders collide at a small grazing
angle. Thus if $v_{12}\approx v_{22}$ the system tends towards an equilibrium
where the small grains in total contain four times more mass than the
boulders. We show in Appendix \ref{s:stability} that any perturbation to the
equilibrium solution will decay on a collisional time-scale, so that
\Eq{eq:rhorat_equi} constitutes a (both linearly and non-linearly) stable
solution to the coagulation-fragmentation problem [Eq.\ (\ref{eq:dn1dt})].

\begin{figure}
  \includegraphics{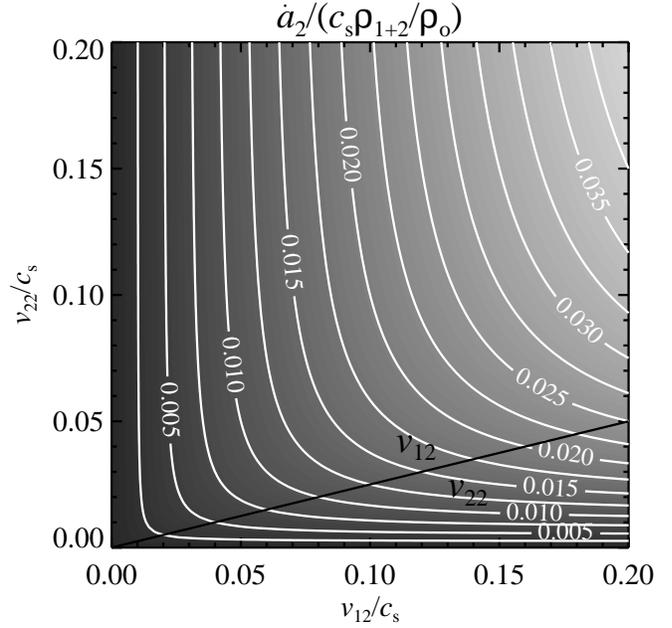}
  \caption{Contour lines of the radius growth of boulders, $\dot{a}_2$, as a
    function of the collision speed between boulders and fragments, $v_{12}$,
    and between boulders and boulders, $v_{22}$. Two regimes are divided by the
    black line: for $v_{22}/v_{12}\ll1/4$ the radius growth depends only on
    $v_{22}$, whereas the radius growth depends only on $v_{12}$ in the limit
    $v_{22}/ v_{12}\gg1/4$. Note that the normalisation of collision speeds
    with sound speed is an arbitrary choice.}
  \label{f:ap2dot_dv12_dv22}
\end{figure}

Inserting \Eq{eq:rhorat_equi} into \Eq{eq:da2dt} yields the evolution of the
boulder radius in the equilibrium state as
\begin{equation}\label{eq:da2dt_equi}
  \frac{\dpa a_2}{\dpa t} =
      \frac{\rho_{1+2}}{\rho_\bullet}
      \frac{v_{22}}{1+4v_{22}/v_{12}} \, .
\end{equation}
Here we have introduced the bulk density of solids $\rho_{1+2}=\rho_1+\rho_2$
which is constant in time in absence of evaporation and condensation
processes. We show in \Fig{f:ap2dot_dv12_dv22} the dependence of the radius
growth on the collision speeds $v_{12}$ and $v_{22}$. It is clear from
\Eq{eq:da2dt_equi} that the radius growth depends only on either $v_{12}$ or
$v_{22}$ in the two limits of $v_{22}/v_{12}$ (i.e.\ zero and infinity). The
dividing line at $v_{22}/ v_{12}=1/4$ is indicated with a black line in
\Fig{f:ap2dot_dv12_dv22}.

\Eq{eq:da2dt_equi} implies that there is a linear growth of the boulder radius
with time, although only under the assumption that $\rho_{1+2}$, $v_{12}$ and
$v_{22}$ are independent of particle size (we shall include the full dependence
of these parameters on the particle size in the numerical simulations presented
in \S\ref{s:simulations} and \S\ref{s:sed}). For typical values of
$\rho_{1+2}/\rho_\bullet=10^{-11}$, $v_{12}=25\,{\rm m\,s^{-1}}$ and
$v_{22}=10\,{\rm m\,s^{-1}}$, relevant in a sedimentary mid-plane layer of
solids at $r=5\,{\rm AU}$ of a moderately turbulent minimum mass solar nebula
model with a turbulent viscosity of $\alpha=10^{-3}$ (see \S\ref{s:simulations}
for a definition of $\alpha$), the growth rate is
$\dot{a}_2=4\times10^{-11}\,{\rm m\,s^{-1}}$, or 1.2 millimeters per year.
Around 2,000 years are then needed to grow from 30 cm, the size for which
radial drift is the fastest, to 3 m in radius, which is so loosely coupled to
the gas that radial drift is no longer a problem. In the absence of collisional
fragmentation, on the other hand, the sweep-up will end after the boulders have
incorporated all the small grains. Considering a fixed number density of
boulders $n_2$, we can write the ratio of the particle radii $a_2$ and $a_2'$
for two different mass densities $\rho_2$ and $\rho_2'$ as
\begin{equation}\label{eq:a2_nofrag}
  \left( \frac{a_2'}{a_2} \right)^3 = \frac{\rho_2'}{\rho_2} \, .
\end{equation}
Setting $\rho_2'=\rho_1+\rho_2$, it is seen from \Eq{eq:a2_nofrag} that if 4/5
of the dust mass is originally in small grains, then the sweep-up of those
grains by the boulders can only lead to a moderate increase of approximately
$70\%$ in the average boulder radius. Only when collisional fragmentation is
included can the boulders grow larger than that, because the reservoir of
grains to sweep up will be continuously replenished.

For the case $v_{12}\approx v_{22}$, \Eq{eq:da2dt_equi} simplifies down to
\begin{equation}\label{eq:da2dt_same}
  \frac{\dpa a_2}{\dpa t} =
      \frac{1}{5} \frac{\rho_{1+2}}{\rho_\bullet} v_{22} \, .
\end{equation}
One can consider yet another special case of \Eq{eq:da2dt_equi} where the
collisions between boulders and grains happen at a much higher speed than the
collisions between boulders and boulders, $v_{12} \gg v_{22}$. This is
relevant if the boulders migrate radially inwards due to a radial pressure
gradient in the gas \citep{Weidenschilling1977}. The drift speed can approach
10\% of the sound speed, easily an order of magnitude higher than the turbulent
gas motions that cause the collisions between boulders. Thus sweep-up works
much more efficiently than fragmentation, and \Eq{eq:da2dt_equi} changes to
\begin{equation}\label{eq:da2dt_drift}
  \frac{\dpa a_2}{\dpa t} =
      \frac{\rho_{1+2}}{\rho_\bullet} v_{22} \, ,
\end{equation}
which is five times faster than \Eq{eq:da2dt_same}. The approximation that
$v_{\rm 22} \ll v_{12}$ may nevertheless be unachievable, even if turbulent
motion is weak, since shape effects will induce differential radial drift even
between equal-mass bodies \citep{Benz2000}.

\subsection{Timescale to reach equilibrium}

\begin{table*}
  \begin{center}
    \begin{tabular}{cccccccccc}
      \hline
        Run & Resolution & Particles & Leaking &
        $\rho_\bullet/\rho_{\rm g}$ &
        $\varSigma_1/\varSigma_{\rm g}$ & $\varSigma_2/\varSigma_{\rm g}$ &
        $a_2/H$ & $\alpha$ & Simulation time\\
        \hline
        A   & $64^3$  & $2.0\times10^6$ & No  & $10^{11}$ & $0.3$  & $0.3$  &
              $10^{-11}$ & $10^{-3}$ & $200 T_{\rm orb}$ \\
        B   & $64^3$  & $2.0\times10^6$ & No  & $10^{11}$ & $0.3$  & $0.3$  &
              $10^{-11}$ & $10^{-2}$ & $200 T_{\rm orb}$ \\
        C   & $64^3$  & $2.5\times10^5$ & Yes & $10^{11}$ & $0.01$ & $0.01$ &
              $10^{-11}$ & $10^{-3}$ & $400 T_{\rm orb}$ \\
        D   & $64^3$  & $2.5\times10^5$ & Yes & $10^{11}$ & $0.01$ & $0.01$ &
              $10^{-11}$ & $10^{-2}$ & $400 T_{\rm orb}$ \\
        E   & $128^3$ & $2.0\times10^6$ & Yes & $10^{11}$ & $0.01$ & $0.01$ &
              $10^{-11}$ & $10^{-3}$ & $200 T_{\rm orb}$ \\
        F   & $128^3$ & $2.0\times10^6$ & Yes & $10^{11}$ & $0.01$ & $0.01$ &
              $10^{-11}$ & $10^{-2}$ & $200 T_{\rm orb}$ \\
        \hline
    \end{tabular}
  \end{center}
  \caption{Simulation parameters. The box size is fixed at $(1.32 H)^3$ in all
    runs. The initial column densities $\varSigma_1$ and $\varSigma_2$ are set
    in runs A and B to mimic the density in a sedimentary mid-plane layer; in
    the simulations with vertical gravity on the boulders (runs C-F), where
    dust fragments can leak freely out of the boulder layer, we set the column
    densities to a more canonical value of $0.01$ for each component.}
  \label{t:parameters}
\end{table*}
Starting from a state where an equal amount of mass is present in dust and in
boulders, the coagulation-fragmentation equilibrium is reached when a
significant fraction of the boulders have undergone collisions (and the
following fragmentation). From \Eq{eq:dn2dt} the time it takes to get to
equilibrium $t_{\rm eq}$ is given by
\begin{equation}\label{eq:teq0}
  \frac{1}{t_{\rm eq}} = n_2 \sigma_{22} v_{22} \, .
\end{equation}
Assuming that the boulders are spherical, the number density $n_2$ and the
collisional cross section $\sigma_{22}$ can be written in terms of the solid
radius $a_2$, yielding
\begin{equation}\label{eq:teq1}
  t_{\rm eq} = \frac{\rho_\bullet a_2}{3 \rho_2 v_{22}} \, .
\end{equation}
When $a_2<(9/4) \lambda$, where $\lambda$ is the mean free path of the gas, the
friction force is in the Epstein regime (see \App{s:dragforce} for a discussion
of the validity of the Epstein regime for the boulders). Here the friction time
can be written as
\begin{equation}\label{eq:tauf_epstein}
  \tau_{\rm f} = \frac{a_2 \rho_\bullet}{c_{\rm s} \rho_{\rm g}} \, ,
\end{equation}
with $\rho_{\rm g}$ denoting the gas density and $c_{\rm s}$ the sound speed.
Inserting this expression for the friction time in \Eq{eq:teq1} yields
\begin{equation}\label{eq:teq_nodim}
  \varOmega_{\rm K} t_{\rm eq} =
      \frac{\varOmega_{\rm K} \tau_{\rm f}}{3 \epsilon_2 v_{22}/c_{\rm
      s}} \, .
\end{equation}
Here $\varOmega_{\rm K}=\varOmega_{\rm K}(r)$ is the Keplerian angular
frequency of the disc, a measure of the local dynamical time-scale at a given
radial location in the disc, and $\epsilon_2$ is the ratio of the bulk
densities of boulders and gas. We define a dimensionless friction time through
the Stokes number ${\rm St}$ as
\begin{equation}
  {\rm St} = \varOmega_{\rm K} \tau_{\rm f} \, .
\end{equation}
Thus the time it takes to reach a coagulation-fragmentation equilibrium is
independent of the actual density of the surrounding gas. For a given Stokes
number, any decrease in the gas density must be balanced by a similar decrease
in the radius of the boulders, leading to an increase in the number density
that balances out the decrease in collisional cross section in \Eq{eq:teq0}.
The presence of the sound speed in \Eq{eq:teq_nodim} also does not affect the
timescale, since the turbulent collision speed $v_{22}$ must scale with the
sound speed as well\footnote{This is strictly not the case in the presence of
magnetic fields, where the local Alfv\'en speed gives a second velocity scale,
but we shall ignore that complication here.}.
\begin{figure}
  \includegraphics{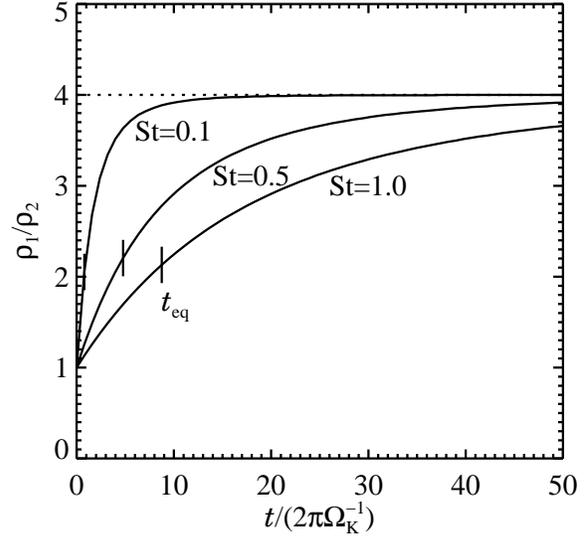}
  \caption{The time evolution of the mass density in dust grains, $\rho_1$,
    relative to that of boulders, $\rho_2$. We have assumed an initial
    dust-to-gas ratio of 0.3 for both dust and boulders, collision speeds of
    $v_{12}=v_{22}=0.02 c_{\rm s}$ and a solid density of
    $\rho_\bullet/\rho_{\rm g}=10^{11}$. The timescale $t_{\rm eq}$ for
    approaching the equilibrium value $\rho_1/\rho_2=4$ (dashed line) is
    indicated with vertical lines.}
  \label{f:rho1rho2_t}
\end{figure}

The time evolution of the mass ratio $\rho_1/\rho_2$ between dust grains and
boulders is shown in \Fig{f:rho1rho2_t}. We have integrated the 0-D
coagulation-fragmentation equations [Eqs.\ (\ref{eq:dn1dt}), (\ref{eq:dn2dt})
and (\ref{eq:da2dt})] with an initial dust-to-gas ratio of 0.3 for both species
(relevant in a mid-plane layer of solids in equilibrium between sedimentation
of turbulent diffusion), collision speeds of $v_{12}= v_{22}=0.02c_{\rm s}$ and
a solid density of $\rho_\bullet/\rho_{\rm g}=10^{11}$. The approach to the
equilibrium state $\rho_1/\rho_2=4$ happens on the equilibrium timescale
$t_{\rm eq}$, given by \Eq{eq:teq1}. For ${\rm St}=0.5$ the equilibrium
timescale is around five orbits, increasing proportional to the Stokes number.
The onset of fragmentation should depend on the radius of the boulders rather
than on the Stokes number, so the critical Stokes number for which collisional
fragmentation gets important depends on the radial location in the disc. Around
the location of Jupiter in a minimum mass nebula a ${\rm St}=1$ particle has a
radius of approximately 30 cm (see \S\ref{s:initial}), for which collisional
fragmentation should already be substantial. Since the
coagulation-fragmentation equilibrium is stable (see \App{s:stability}), the
state will stay at the equilibrium once it is reached.

It is not strictly necessary to be near the equilibrium value of
$\rho_1/\rho_2$ to have efficient growth of the boulders by sweep-up. The
radius of the boulders grows proportionally to $\rho_1$, according to
\Eq{eq:da2dt}, so just maintaining the reservoir of grains is already an
achievement of the collisional fragmentation. As $\rho_1$ increases with time,
the boulders will grow faster and faster until finally reaching the growth
speed given by \Eq{eq:da2dt_equi}.

\section{Simulation set up}
\label{s:simulations}

Next we will solve the coagulation-fragmentation equations numerically in a
three-dimensional time-dependent turbulent flow. In this section we describe
the dynamical equations and the adopted protoplanetary disc model. We let
turbulence arise through the magnetorotational instability
\citep{BalbusHawley1991} which operates when the gas is sufficiently ionised
\citep{Gammie1996,Semenov+etal2004}. Typically the most unstable wavelength of
the magnetorotational instability is around one gas scale height, with a
subsequent energy cascade to smaller scales that approximately obeys a
Kolmogorov-law \citep{Hawley+etal1995}. The saturated state of the
magnetorotational instability (which we will refer to as magnetorotational
turbulence) is characterised by an outwards transport of angular momentum
through positive Reynolds and Maxwell stresses. In shearing box simulations the
measured $\alpha$-value ranges from $10^{-3}$ (with zero net flux field) to
above 0.1 \citep[for $\beta=P_{\rm gas}/P_{\rm mag}=400$ net vertical field,
see][]{Hawley+etal1995}. Numerically, magnetorotational turbulence has the
great advantage that it is relatively easy to produce and sustain in local box
simulations for hundreds of disc rotation periods
\citep{Brandenburg+etal1995,Hawley+etal1995}.

\cite{Gullbring+etal1998} and more recently \cite{Sicilia-Aguilar+etal2004}
measured the accretion luminosities of T Tauri stars and translated the
measurements into mass accretion rate $\dot{M}$. Typical estimated values of
the mass accretion rate lie in the interval $\dot{M}=10^{-9\ldots-7}{\rm
M}_\odot\,{\rm yr^{-1}}$. Coupling the mass accretion rate with a disc model
yields the turbulent viscosity of the disc, $\nu_{\rm t}$, through the relation
\citep{Pringle1981}
\begin{equation}
  \nu_{\rm t} = (3 \pi)^{-1} \frac{\dot{M}}{\varSigma} \, ,
\end{equation}
where $\varSigma$ is the column density of gas and solids. Making use of the
non-dimensionalisation with sound speed $c_{\rm s}$ and angular frequency
$\varOmega_{\rm K}$ of \cite{ShakuraSunyaev1973}, $\nu_{\rm t}=\alpha c_{\rm
s}^2 \varOmega_{\rm K}^{-1}$, we obtain the $\alpha$-value of the disc through
\begin{equation}
  \alpha = (3 \pi)^{-1} \frac{\dot{M}}{\varSigma}
  \frac{\varOmega_{\rm K}}{c_{\rm s}^2} \, .
\end{equation}
For the minimum mass solar nebula $\alpha=10^{-4\ldots-2}$ from typical mass
accretion rates. The turbulent viscosity can be approximated as
\begin{equation}\label{eq:nut}
  \nu_{\rm t} = \tau_{\rm eddy} u_{\rm rms}^2 \, ,
\end{equation}
where $\tau_{\rm eddy}$ is the eddy turn over time and $u_{\rm rms}$ is the
turbulent rms speed. Assuming $\tau_{\rm eddy}\approx\varOmega_{\rm K}^{-1}$,
due to the dominating effect of the Coriolis force at large scales
\citep{Weidenschilling1984}, one obtains $\alpha=(u_{\rm rms}/c_{\rm s})^2$.
However one must be careful when translating $\alpha$ into $u_{\rm rms}$ this
way, since $\alpha$ normally refers to the turbulence's ability to diffuse the
main Keplerian differential rotation, and instabilities that are not Keplerian
shear instabilities are often associated with a negative $\alpha$-value
\citep[such as convection or streaming instability,
see][]{RyuGoodman1992,YoudinGoodman2005}. For magnetorotational turbulence,
\Eq{eq:nut} nevertheless holds relatively well (see \Tab{t:results}).

In this paper we shall focus on two values for the viscosity which we believe
are most relevant (based on the observed accretion rates): low viscosity with
$\alpha=10^{-3}$ (arising in zero net flux simulations) and high viscosity with
$\alpha=10^{-2}$ (the result of simulations with a weak $\beta=20000$ vertical
magnetic field). We use the Pencil Code \citep{Brandenburg2003} to solve the
equations of ideal magnetohydrodynamics, as described in detail in
\cite{JohansenKlahrHenning2006}. For simplicity we ignore vertical
stratification of the gas and model a local corotating shearing box with side
lengths $1.32 H$, where $H=c_{\rm s}/\varOmega_{\rm K}$ is the scale height of
the gas. Our coordinate frame is oriented in such a way that the $x$-axis
points outwards along the radial direction, the $y$-axis points along the main
Keplerian flow, while the $z$-axis points vertically out of the disc in the
direction of the Keplerian frequency vector $\vec{\varOmega}_{\rm K}$. The
simulation parameters are written in \Tab{t:parameters}. The initial condition
for the solids is explained in \S\ref{s:initial}.

\subsection{Sweep-up}

Boulders are treated as individual particles, each with a unique position and
velocity vector. The boulders feel a drag force from the gas, described in
detail in \App{s:dragforce}, but for simplicity we assume that gas feels no
drag from the boulder component. Each particle represents a huge number of
actual boulders, hence we refer to them as superparticles. The dust component
is treated as a passive scalar: the velocity field is set equal to that of the
gas, so that only a continuity equation must be solved, but with additional
source and sink terms (due to destruction of boulders and sweep-up) as
described below.

The superparticles are given two internal degrees of freedom -- the number
density of actual boulders inside each particle $\tilde{n}_i$ and the average
radius $a_i$ of the constituent boulders. A superparticle is allowed to change
the radius of its boulders, $a_i$, by sweeping up dust grains. The dynamical
equation for $a_i$ is
\begin{equation}
  \frac{\dpa a_i}{\dpa t} = \frac{\epsilon_1 \rho_{\rm g}}{4 \rho_\bullet} v_{0i}
  \, .
\end{equation}
Here $\epsilon_1$ is the dust-to-gas ratio of the dust grains, $\rho_{\rm g}$
is the density of the gas, $\rho_\bullet$ is the material density of the
solids, and $v_{0i}$ is the relative speed between superparticle $i$ and the
gas in its grid cell (the dust grains are so coupled to the gas that this is
the same as the collision speed between boulders and grains). The density of
the dust grains is depleted at the same time according to the evolution equation
\begin{equation}
  \frac{\dpa \epsilon_1}{\dpa t} =
      -\tilde{n}_i \pi a_i^2 v_{0i} \epsilon_1 \, ,
\end{equation}
with $\tilde{n}_i$ denoting the number density of boulders in the superparticle
$i$. This number does not change in a sweep-up process.

\subsection{Collisions between boulders}

We identify all superparticles that reside in the same grid cell as colliding.
Collisions between boulders in the superparticles $i$ and $j$ happen at the rate
\begin{equation}
  \dot{c}_{ij}=\sigma_{ij} \tilde{n}_i \tilde{n}_j v_{ij} \, ,
\end{equation}
where $\sigma_{ij}$ is the collisional cross section and $v_{ij}$ is the
collision speed. We assume spherical particles with $\sigma_{ij}=\pi
(a_i+a_j)^2$.

Boulder collisions are assumed to always lead to total destruction of the
colliding bodies. The internal number densities of the colliding superparticles
$i$ and $j$ change as
\begin{eqnarray}
  \frac{\dpa \tilde{n}_i}{\dpa t} &=& -\dot{c}_{ij} \, , \\
  \frac{\dpa \tilde{n}_j}{\dpa t} &=& -\dot{c}_{ij} \, ,
\end{eqnarray}
which has to be considered for all combinations of $i$ and $j$ in each grid
cell. The total mass that is lost in destructive collisions is subsequently
transferred to the dust component. Here the dust-to-gas ratio increases as
\begin{eqnarray}
  \rho_{\rm g} \frac{\dpa \epsilon_1}{\dpa t} =
      \frac{4}{3} \pi \rho_\bullet (a_i^3+a_j^3) \dot{c}_{ij} \, .
\end{eqnarray}
One can think of many improvements to these simplified
coagulation-fragmentation evolution equations, but we believe that it is
enlightening to consider the most simple dynamical equation system that
displays coagulation-fragmentation growth. We shall also compare our results to
the advanced models presented in BDH where a size distribution of solids is
considered and where the fragmentation model is much more advanced and show
that our results are in relatively good agreement with this more advanced model.

One may suspect that the sedimentary mid-plane layer could be dense enough to
have a gravitational influence on the produced fragments. The gravitational
acceleration of a homogeneous, infinitely extended mid-plane layer with density
profile $\rho_{\rm p}(z)=[\varSigma_{\rm p}/(\sqrt{2\pi} H_{\rm p})]
\exp[-z^2/(2 H_{\rm p}^2)]$ is
\begin{equation}
  g_z=-2\pi G \varSigma_{\rm p} {\rm erf}\left( \frac{z}{\sqrt{2} H_{\rm p}}
  \right) \, .
\end{equation}
This acceleration is 3-4 orders of magnitude smaller than the vertical gravity
from the central star ($g_z=-\varOmega_{\rm K}^2 z$). Thus the self-gravity of
the sedimentary mid-plane layer can have no influence on the escape rate of
fragments and we shall ignore the effect of self-gravity in this paper.

\subsection{Radial drift}

The accretion process causes the gas pressure in protoplanetary discs to fall
with radial distance from the young star. We can write the global radial
pressure gradient acceleration as
\begin{equation}
  -\frac{1}{\rho_{\rm g}} \frac{\dpa P}{\dpa r} =
  -\frac{c_{\rm s}^2}{r} \frac{\dpa \ln P}{\dpa \ln r} =
  -\varOmega_{\rm K} c_{\rm s} \frac{H}{r} \frac{\dpa \ln P}{\dpa \ln r} \, .
\end{equation}
Here $H/r$ is the disc aspect ratio and $c_{\rm s}$ is the sound speed. The
balance between Coriolis force and global pressure gradient gives the gas
orbital velocity relative to the Keplerian motion as
\begin{equation}
  \frac{u_y^{\rm (subK)}}{c_{\rm s}} = \frac{1}{2} \frac{H}{r} \frac{\dpa \ln P}{\dpa \ln r}
  \, .
\end{equation}
It is common to define the pressure gradient parameter $\eta$ as
\citep{Nakagawa+etal1986}
\begin{equation}
  \eta = -\frac{1}{2} \left( \frac{H}{r} \right)^2 
         \frac{\dpa \ln P}{\dpa \ln r} \, ,
\end{equation}
giving $u_y^{\rm (subK)}=-\eta v_{\rm K}$. Here $v_{\rm K}=\varOmega_{\rm K} r$
is the Keplerian orbital speed.

The boulders do not feel the global pressure gradient and would orbit with the
local Keplerian speed in absence of gas. The head wind of the slower moving gas
drains the boulders of angular momentum and imposes a flux of boulders towards
smaller $r$. The equilibrium radial drift velocity of the boulders is given by
\citep{Weidenschilling1977,YoudinJohansen2007}
\begin{equation}
  v_x = -\frac{2 \eta v_{\rm K}}{\varOmega_{\rm K} \tau_{\rm f}+(\varOmega_{\rm
  K} \tau_{\rm f})^{-1}} \, .
\end{equation}
Typical values of $\eta v_{\rm K}$ lie between $0.02 c_{\rm s}$ and $0.1 c_{\rm
s}$ \citep{Nakagawa+etal1986}. In this paper we assume that $v_x=-0.05 c_{\rm
s}$ for marginally coupled boulders with $\varOmega_{\rm K} \tau_{\rm f}=1$. As
in \cite{JohansenKlahrHenning2006} we apply the global pressure gradient force
directly on the boulders instead of on the gas. This simplified treatment of
radial drift is valid as long as the drag force from the boulders on the gas is
ignored.

\subsection{Units and initial condition}
\label{s:initial}

We adopt a dimensionless unit system in our corotating box by setting the sound
speed $c_{\rm s}=1$, Keplerian frequency $\varOmega_{\rm K}=1$ and mid-plane
gas density $\rho_{\rm g}(z=0)=1$ (for computational simplicity we ignore gas
stratification, so the gas density is approximately one everywhere in the
box). Thus gas and particle velocities are measured in units of the sound
speed, while length and particle radius is in units of gas scale heights
$H=c_{\rm s}/\varOmega_{\rm K}=1$. For the magnetic fields we set the vacuum
permeability $\mu_0=1$. We stress that the role of magnetic fields in our model
is to tap into the Keplerian motion and release gravitational energy as
turbulent kinetic energy, which is transferred from the gas to the solids by
drag forces, but that dust grains and boulders are otherwise unaffected by
magnetic fields.

Physical parameters for the different runs are written in Table
\ref{t:parameters}. The initial solids-to-gas ratio of boulders and fragments
is set to $0.3$ for both species in the two runs where dust fragments are not
allowed to diffuse out of the boulder layer (runs A-B), to mimic the density in
a sedimentary mid-plane layer\footnote{Actually it is not very realistic to
have such a high density of small dust grains in a sedimentary mid-plane layer
to begin with. Therefore we also ran variations of runs A-B with all the mass
initially in the boulder component, but found essentially the same results.} ,
whereas the ratio between solids and gas column densities is set to the more
canonical $0.01$ in the simulations with vertical gravity acting on the
boulders (runs C-F). The dynamical equations of coagulation-fragmentation
furthermore depend on internal properties of the solids: the radius of the
boulders $a_2$ and the material density of the solids $\rho_\bullet$. These
must be defined in code units. We initially set $a_2=10^{-11}$ and
$\rho_\bullet=10^{11}$, giving an initial Stokes number of unity through ${\rm
St}_2=\varOmega_{\rm K} \tau_2=(a_2/H)(\rho_\bullet/\rho_{\rm g})=1$.

At $r=5\,{\rm AU}$ in the minimum mass solar nebula with sound speed $c_{\rm
s}=5\times10^4\,{\rm cm\,s^{-1}}$, Keplerian frequency $\varOmega_{\rm
K}=1.7\times10^{-8}\,{\rm s^{-1}}$ and gas column density $\varSigma_{\rm
g}=150\,{\rm g\,cm^{-2}}$ our dimensionless model corresponds to a gas scale
height of $H=3\times10^{12}\,{\rm cm}$, a mid-plane gas density of $\rho_{\rm
g}(z=0)=2 \times 10^{-11}\,{\rm g\,cm^{-3}}$, a material density of
$\rho_\bullet=2\,{\rm g\,cm^{-3}}$ and an initial boulder radius of
$a_2=30\,{\rm cm}$.
\begin{figure*}
  \begin{center}
    \includegraphics[width=0.3\linewidth]{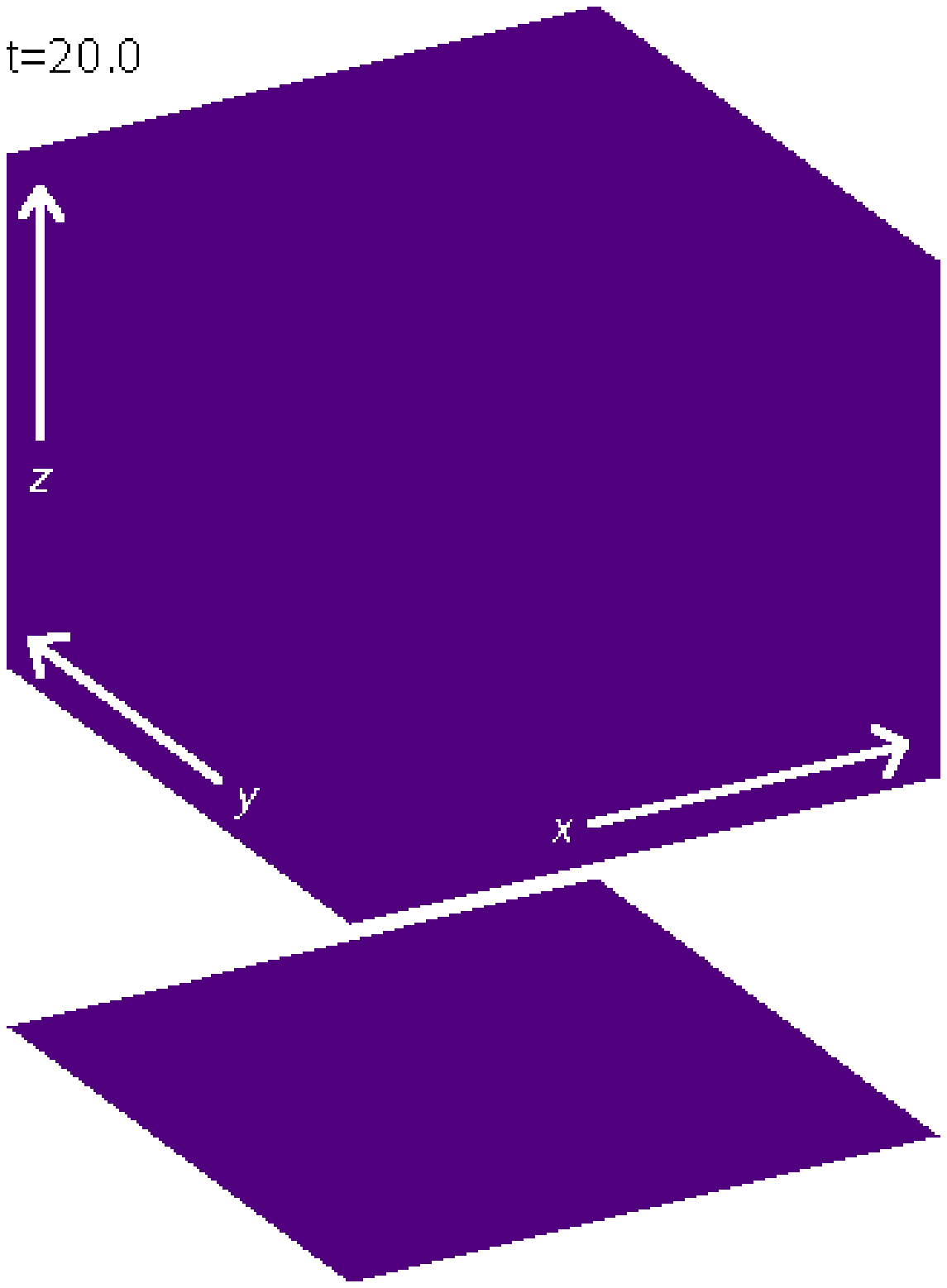}
    \includegraphics[width=0.3\linewidth]{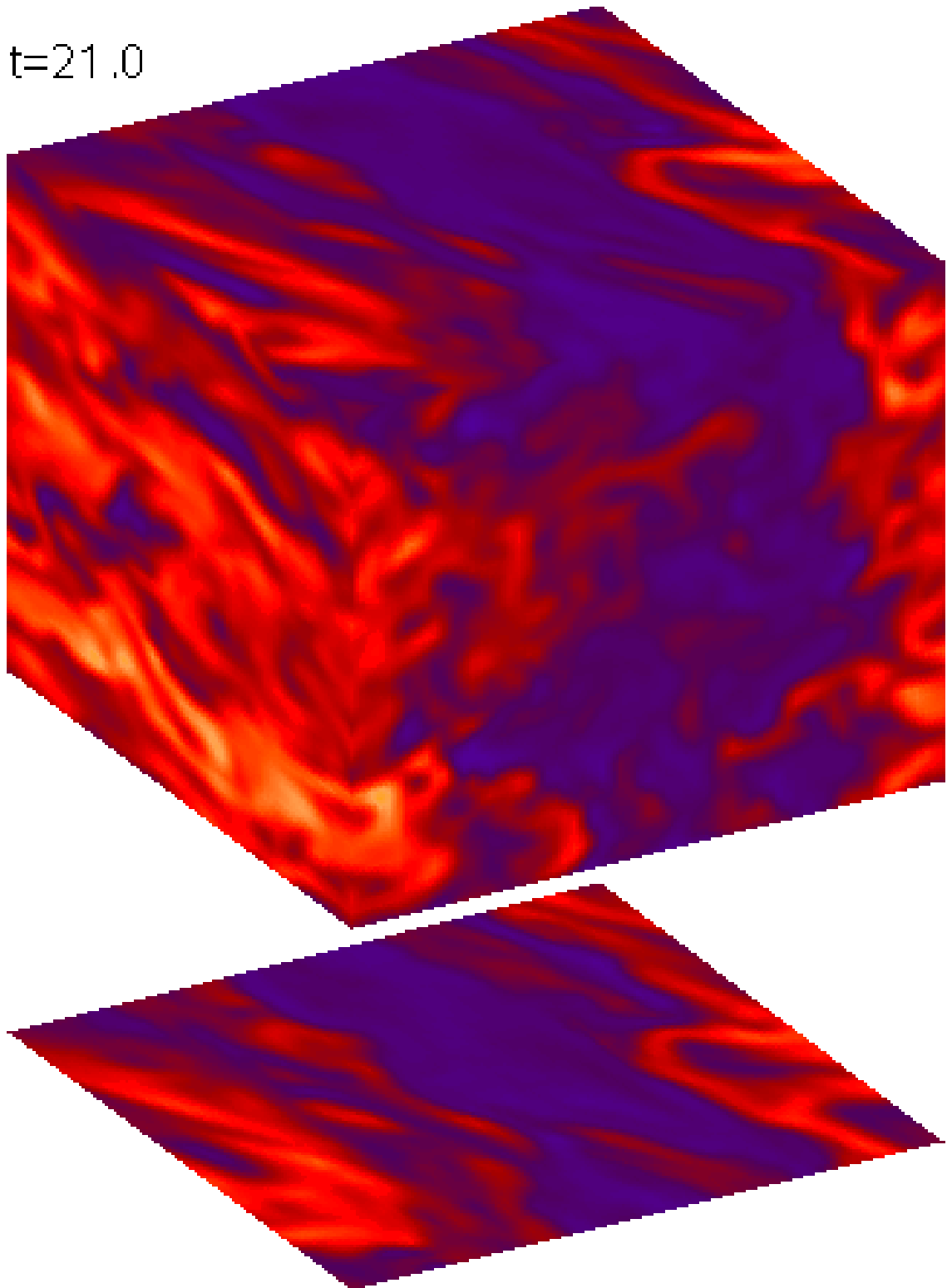}
    \includegraphics[width=0.3\linewidth]{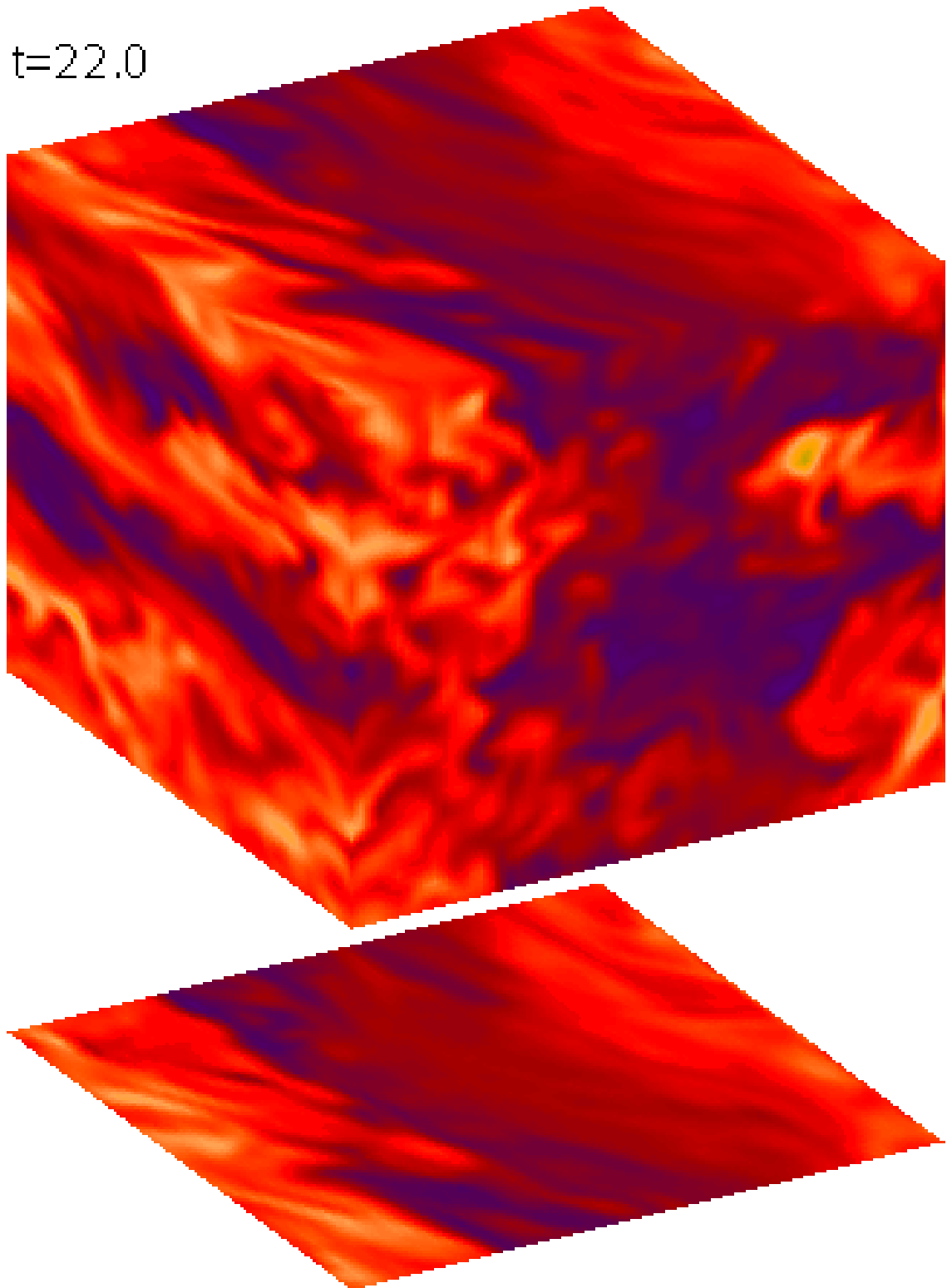}
    \includegraphics[width=0.3\linewidth]{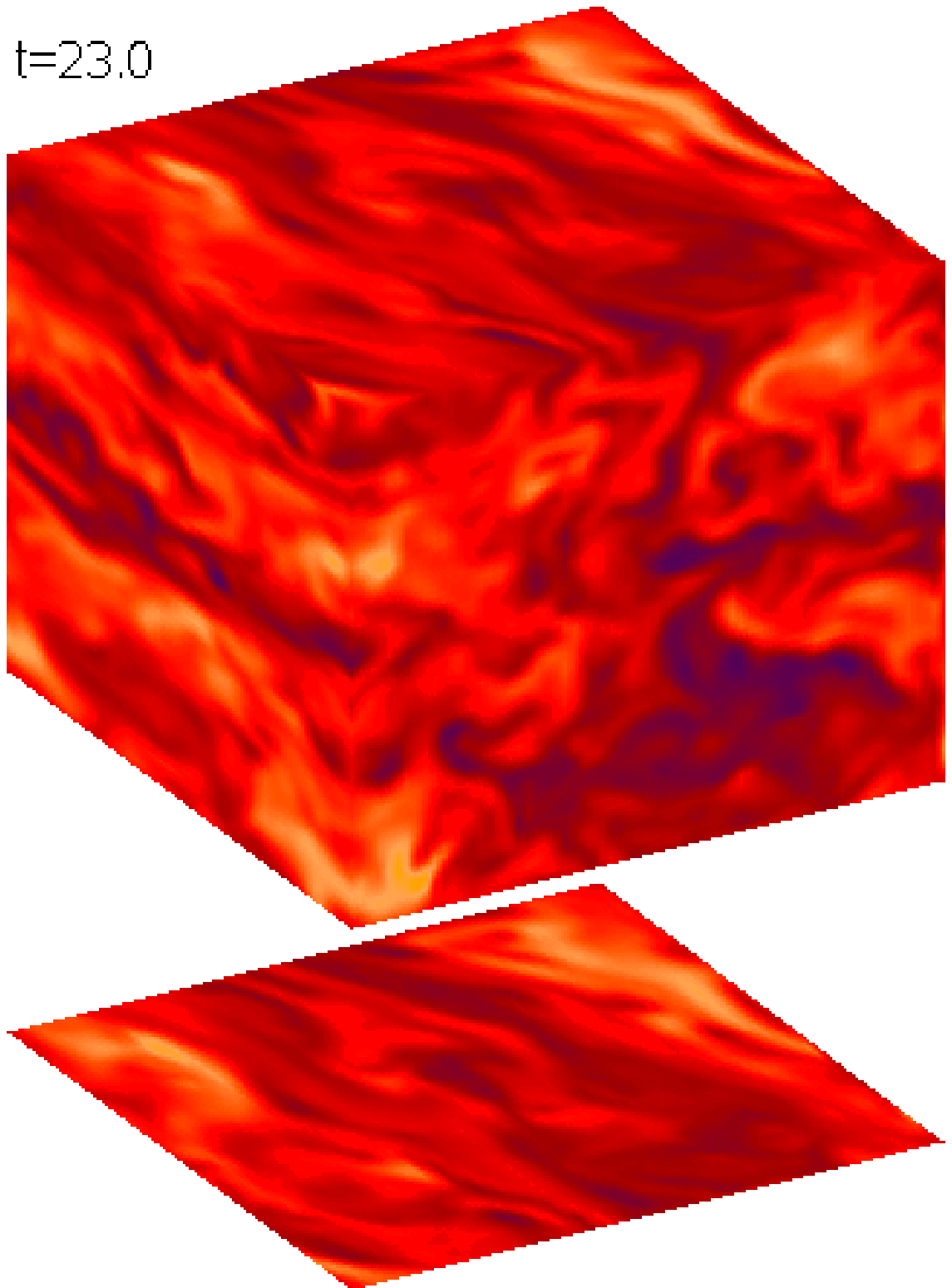}
    \includegraphics[width=0.3\linewidth]{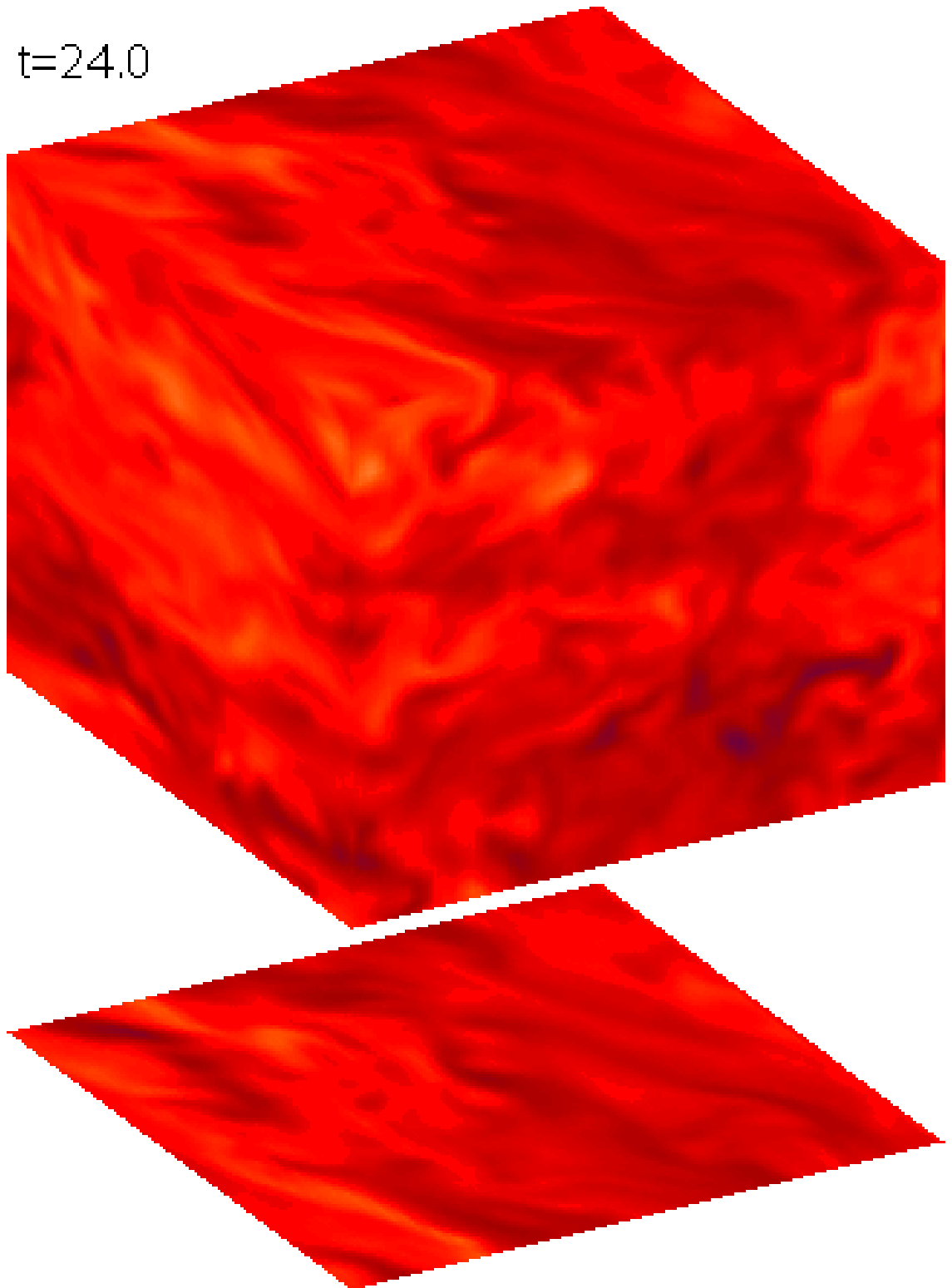}
    \includegraphics[width=0.3\linewidth]{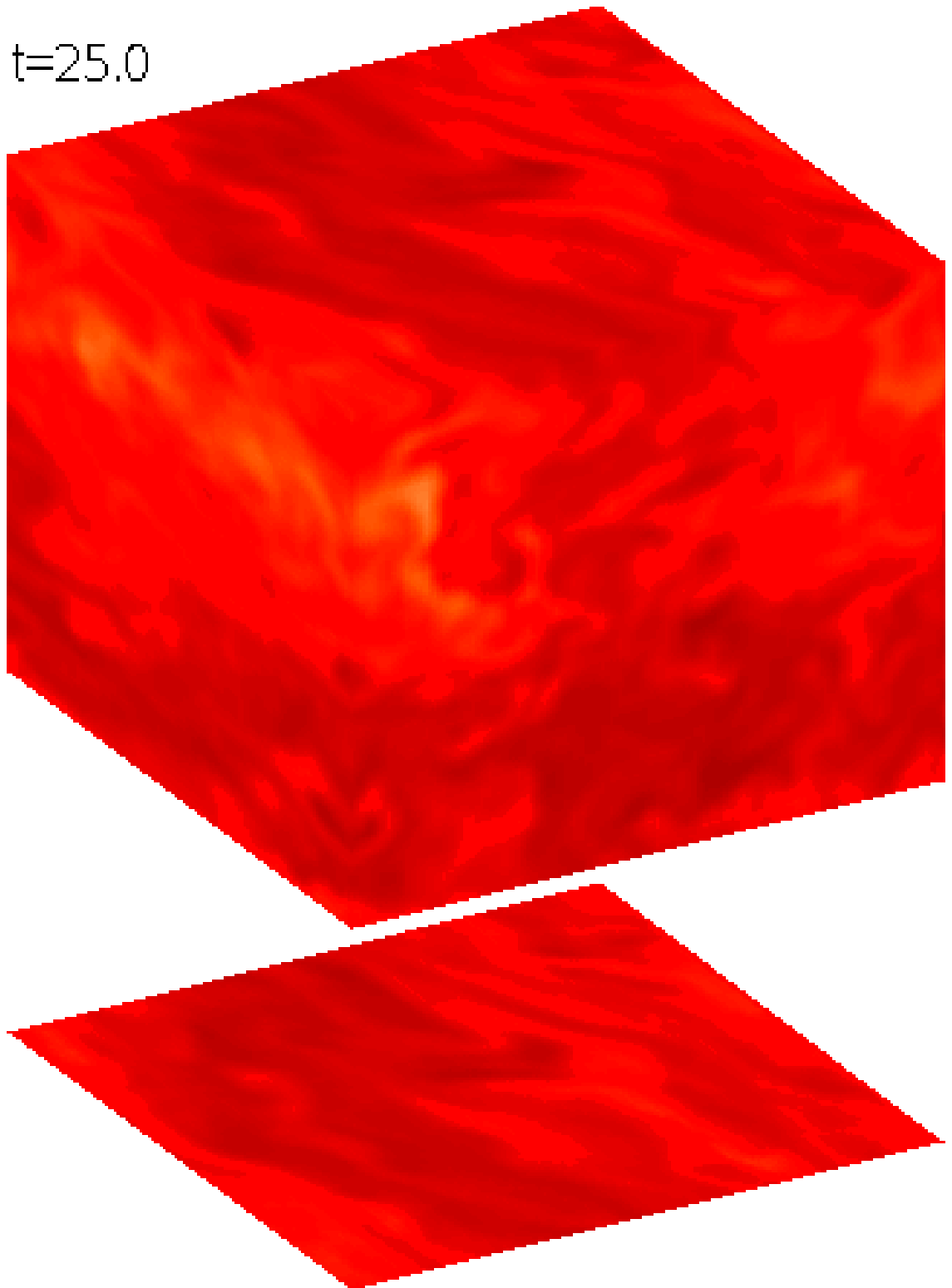}
  \end{center}
  \caption{Time series of the concentration of small dust grains for the run
    with $\alpha=0.01$ and no vertical gravity on the boulder component (run
    B). The boxes are oriented with the radial $x$-axis to the right and
    slightly up, the azimuthal $y$-axis to the left and up, and the vertical
    $z$-axis directly up. The time $t$ is given in local orbital periods --
    coagulation and collisional fragmentation are turned on after 20 orbits to
    avoid any effect of the initial condition on the evolution of the solids.
    Fragments immediately form in catastrophic collisions between boulders.
    Even though the boulder component is clumpy, evident from the inhomogeneous
    initial production of fragments, the dust fragments eventually spread out
    evenly through the box. This homogenisation is an effect of the turbulent
    diffusion time-scale being much shorter than the collisional time-scale.}
  \label{f:fragmentation_panels}
\end{figure*}

\section{Results neglecting leaking}\label{s:nosed}

We first treat models where the dust fragments are not allowed to leave the
boulder layer (runs A and B in \Tab{t:parameters}) to test the validity of the
analytical model described in \S\ref{s:analytics}. We mimic the physical
conditions in a sedimentary mid-plane layer of boulders by setting the
densities of boulders and dust grains artificially high (both components are
given a solids-to-gas ratio of 0.3). In the following sections,
\S\ref{s:diffusion}--\ref{s:sed}, we treat the more realistic case where
boulders lie in a thin layer around the mid-plane and where the dust fragments
can leave this layer due to turbulent diffusion. {\it We caution the reader
already now that allowing the dust fragments to escape from the sedimentary
mid-plane layer will make prospects to cross the meter barrier by
coagulation-fragmentation much more negative than they appear in this section.}

In \Fig{f:fragmentation_panels} we show snapshots of the density of fragments
(relative to the local gas density) as a function of time. The turbulence has
been given 20 orbits to develop before the sweep-up and fragmentation terms are
turned on, to avoid the initial condition having any influence on the results.
After one orbit (at $t=21T_{\rm orb}$) fragments have formed in collisions
between boulders. The fragments are continuously mixed by the turbulence, and
after a few orbits the disc reaches a state where the fragments are very
well-mixed with the gas. This state is preferred even though the boulder
component is not homogeneous, because the collision time-scale of boulders is
much longer than the diffusion time-scale. Considering an overdense region of
size $\Delta$, the time-scale for fragments to diffuse out of this region is
$t_{\rm diff} = \Delta^2/D_{\rm t}$, where $D_{\rm t}$ is the diffusion
coefficient, while the collisional time-scale is $t_{\rm coll} = m_2/(\rho_2
\sigma_{22} v_{22})$. The ratio of the two time-scales is
\begin{equation}
  \frac{t_{\rm diff}}{t_{\rm coll}} = 
  \frac{3 (\Delta/H)^2/\delta}{\varOmega_{\rm K} \tau_{\rm f} (
  v_{22}/c_{\rm s})^{-1} (\rho_2/\rho_{\rm g})^{-1}} \, .
\end{equation}
Here we have used the parametrisation $D_{\rm t}=\delta c_{\rm s}^2
\varOmega_{\rm K}^{-1}$ and the Epstein friction time $\tau_{\rm f}=a_2
\rho_\bullet/(c_{\rm s} \rho_{\rm g})$. Using typical values for run B,
$\Delta/H=0.1$, $\delta=10^{-2}$, $\varOmega_{\rm K} \tau_{\rm f}=1$,
$v_{22}/c_{\rm s}=0.07$ and $\rho_2/\rho_{\rm g}=1$ in the overdense regions,
yields $t_{\rm diff}/t_{\rm coll} \approx 0.2$. Thus the fragments have plenty
of time to escape the overdense regions before they are swept up by the
boulders there, leading to an almost homogeneous spatial distribution of dust
fragments at $t=25 T_{\rm orb}$ in \Fig{f:fragmentation_panels}.
\begin{figure*}
  \begin{center}
    \includegraphics[width=\linewidth]{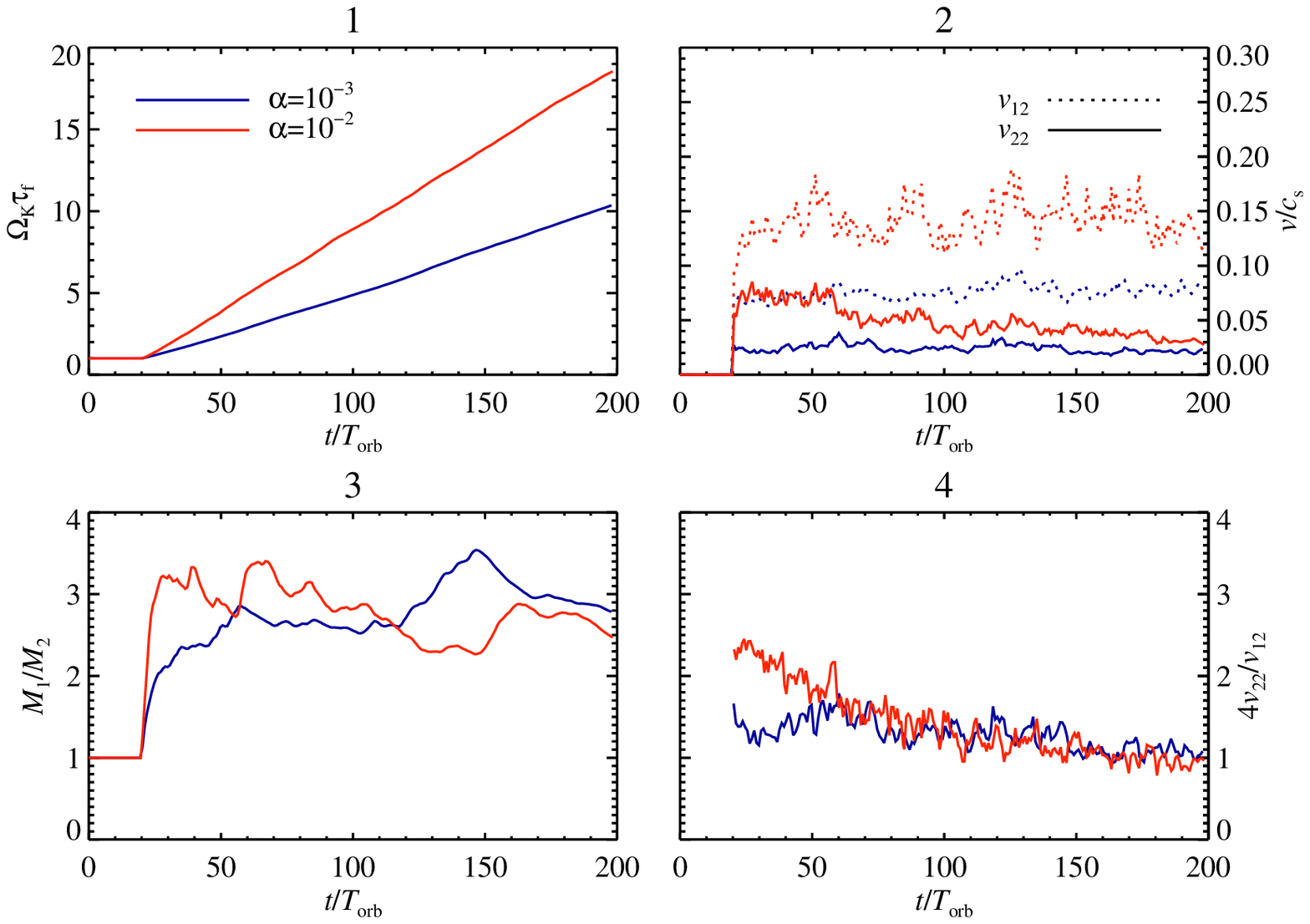}
  \end{center}
  \caption{Growth and fragmentation of boulders as a function of time $t$,
    measured in orbits, for two different strengths of the turbulence. The
    first panel shows the friction time of the boulders: growth from the
    initial $\varOmega_{\rm K}\tau_{\rm f}=1$ to beyond $\varOmega_{\rm
    K}\tau_{\rm f}=10$ occurs readily for both weak and strong turbulence,
    although the stronger turbulence helps growth by producing more fragments
    (note that the Epstein drag law is formally not valid beyond
    $\varOmega_{\rm K} \tau_{\rm f}\approx7$, see discussion in
    \App{s:dragforce}, but since this state can already be considered as having
    crossed the meter barrier, we have ignored the complications of switching
    to a Stokes drag law). The second panel shows the relative speeds between
    boulders and fragments ($v_{12}$) and between boulders and boulders
    ($v_{22}$). The boulder collision speed decreases with time as the
    particles grow and decouple from the gas, whereas the relative speed
    between grains and boulders is set primarily by the turbulent motion of the
    dust grains and thus stays approximately constant. The total mass of the
    fragments, $M_1$, reaches 2-3 times the mass of boulders, $M_2$, in the
    equilibrium state (panel 3), somewhat higher than the analytical
    expectation (panel 4) based on \Eq{eq:rhorat_equi}, but that is likely due
    to the fact that the boulder density is not homogeneous: fragments are
    created where the boulder density is high, but quickly escape these regions
    by turbulent diffusion, leading to an increase in the total amount of dust.}
  \label{f:dvp_t}
\end{figure*}

In \Fig{f:dvp_t} we show the evolution of the friction time of boulders (first
panel), the relative speeds between fragments and boulders $v_{12}$ and between
boulders and boulders $v_{22}$ (second panel), the ratio of total fragment mass
to total boulder mass $M_1/M_2$ (third panel), and the analytical expectation
for $M_1/M_2$ (fourth panel) based on \Eq{eq:rhorat_equi}. The growth of
boulders by sweep-up is very efficient and allows growth to $\varOmega_{\rm K}
\tau_{\rm f}\gtrsim10$ in 100-200 orbits, around the same time-scale as the
radial drift. At this size (approximately 10 meters) radial drift is
insignificant and the boulders are no longer in any risk of being lost to the
inner part of the disc. The variation in particle radius at a given time (not
shown in \Fig{f:dvp_t}) is generally within 10\% of the average value, due to
the strong coupling between different regions of the flow by turbulent
diffusion.

Average values of collision speeds and boulder growth rates for all the
simulations are written in \Tab{t:results}. The growth rate is almost twice as
high in run B (with $\alpha=10^{-2}$) than in run A ($\alpha=10^{-3}$) because
of the higher collision speeds in the strongly turbulent case.
\begin{table*}
  \begin{center}
    \begin{tabular}{ccccccccc}
      \hline
        Run & Res & Leaking & $\alpha$ &
              $v_{12}$ & $v_{22}$ &
              $\varSigma_1/\varSigma_2$ & $\dot{a}_2$ & $t_{\rm recyc}$ \\
        \hline
        A & $64^3$  &  No & $10^{-3}$ &
            $0.067$ & $0.023$ & $2.82$   & $0.057$  & $3.1$   \\
        B & $64^3$  &  No & $10^{-2}$ &
            $0.12$  & $0.071$ & $2.54$   & $0.089$   & $2.5$   \\
        C & $64^3$  & Yes & $10^{-3}$ &
            $0.066$ & $0.014$ & $94.0^*$ & $0.0036$ & $803.5$ \\
        D & $64^3$  & Yes & $10^{-2}$ &
            $0.12$  & $0.054$ & $57.6$   & $0.0064$  & $306.4$ \\
        E & $128^3$ & Yes & $10^{-3}$ &
            $0.063$ & $0.010$ & $58.0^*$ & $0.0029$ & $640.5$ \\
        F & $128^3$ & Yes & $10^{-2}$ &
            $0.16$  & $0.075$ & $40.0^*$ & $0.0068$  & $238.6$ \\
        \hline
    \end{tabular}
  \end{center}
  \caption{Results. The collision speeds $v_{12}$ and $v_{22}$ are averaged
    over orbits 20--30, while the column density ratio
    $\varSigma_1/\varSigma_2$ and boulder radius growth rate $\dot{a}_2$ (given
    here in Stokes number per orbit) are averaged over the last 10 orbits in
    the simulation. The recycling time-scale $t_{\rm recyc}$ is calculated from
    \Eqs{eq:trecyc}{eq:trecycsed}. Measurements marked with * had not yet
    saturated at the end of the simulation.}
  \label{t:results}
\end{table*}

\subsection{Clumping}
\label{s:clumping}

There is a discrepancy of around a factor $2$--$3$ between the analytical
expectation of $M_1/M_2$ in \Fig{f:dvp_t} (panel 4) and the measured value
(panel 3). This is likely due to the fact that the boulder layer is not
homogeneous because the particles are concentrated in high pressure regions of
the gas \citep{JohansenKlahrHenning2006}. Fragments are primarily produced in
the overdense regions, but they quickly mix in with the gas (which is
approximately isodense since the turbulence is subsonic). Thus the high density
regions must produce more fragments to keep up with the turbulent diffusion,
and this increases the overall ratio of fragments to boulders.

One could have thought that local overdensities in the boulder layer would
allow for enhanced collisional fragmentation and thus faster growth out of the
radial drift regime. This appealing picture nevertheless turns out to be
incorrect. To see the effect of clumping we imagine collecting the material
from $N$ grid points into one single grid point, or equivalently to take the
boulders from a volume $V$ and press them together in the volume $V/N$. The
total mass density is $N \langle \rho_{1+2} \rangle$. The single grid point
that contains boulders must fulfil \Eq{eq:rhorat_equi} in order to be in
equilibrium. This leads to the equation system
\begin{eqnarray}
  N \rho_1 + \rho_2 &=& N \langle \rho_{1+2} \rangle \, , \label{eq:simple1} \\
  \rho_1 - 4 \zeta \rho_2 &=& 0 \, , \label{eq:simple2}
\end{eqnarray}
where we define $\zeta\equiv v_{22}/v_{12}$ and assume that the diffusion
time-scale is much shorter than the collisional time-scale so that $\rho_1$
will be constant among all grid cells. The solution to the algebraic equation
system is
\begin{eqnarray}
  \rho_1 &=& \frac{N}{N+1/(4 \zeta)} \langle \rho_{1+2} \rangle \, , \\
  \rho_2 &=& \frac{N}{4 \zeta N+1} \langle \rho_{1+2} \rangle \, .
\end{eqnarray}
For $N=1$, corresponding to no clumping, we recover the usual $\rho_1=(4/5)
\langle \rho_{1+2} \rangle$ and $\rho_2=(1/5) \langle \rho_{1+2} \rangle$ for
$\zeta=1$. For $N\rightarrow\infty$ the expressions tend towards
$\rho_1=\langle \rho_{1+2} \rangle$ and $\rho_2=[1/(4\zeta)] \langle \rho_{1+2}
\rangle$. Thus clumping has little or no effect on the equilibrium density of
dust fragments, which is the crucial parameter that determines radius growth.
The overdense regions must produce enough dust not only to feed its own zone,
but also to fill up the regions that contain no boulders, as any gradients in
the dust density will be quickly [instantaneously actually, in the simplified
model presented in Eqs.\ (\ref{eq:simple1})--(\ref{eq:simple2})] evened out by
diffusion. All together clumping leads maximally to a 25\% increase in radius
growth, but at the cost of reducing the total amount of boulders proportionally
to the degree of clumping. We investigate the role of turbulent diffusion
further in \S\ref{s:diffusion}.

\subsection{Recycling time-scale}
\label{s:recyc}

Even though a balance between sweep-up and collisional fragmentation arises, so
that the number density of fragments $n_1$ stays approximately constant in
time, there is a significant flux of dust grains through the boulder
component. The evolution equation for $n_1$ [\Eq{eq:dn1dt}] consists of two
terms that balance out in the equilibrium. One can rewrite the evolution
equation in terms of the time-scale for sweep-up $t_{\rm sweep}$ and the
time-scale for replenishment of dust fragments by boulder collisions $t_{\rm
replenish}$,
\begin{equation}
  \frac{1}{\rho_1} \frac{\dpa \rho_1}{\dpa t} = -\frac{1}{t_{\rm sweep}} +
  \frac{1}{t_{\rm replenish}} \, .
\end{equation}
In equilibrium the sweep-up and replenishment time-scales are equal. The
sweep-up time-scale is
\begin{equation}
  t_{\rm sweep} = \frac{1}{n_2 \sigma_{12} v_{12}} \, .
\end{equation}
Inserting the equilibrium solution $\rho_1/\rho_2=4 v_{22}/ v_{12}$ from
\Eq{eq:rhorat_equi} gives
\begin{equation}
  t_{\rm sweep} = \frac{1+4 v_{22}/v_{12}}{v_{12}}
  \frac{a_2 \rho_\bullet}{3 \rho_{1+2}}
\end{equation}
when assuming spherical grains with material density $\rho_\bullet$. One can
simplify the equation further by inserting the Epstein regime friction time
from \Eq{eq:tauf_epstein}, yielding
\begin{equation}\label{eq:trecyc}
  \frac{t_{\rm sweep}}{\tau_{\rm f}} = 
  \frac{1+4 v_{22}/v_{12}}{3} \left( \frac{\rho_{1+2}}{\rho_{\rm
  g}} \right)^{-1} \left( \frac{v_{12}}{c_{\rm s}} \right)^{-1} \, .
\end{equation}
This is the time-scale upon which the boulders would empty the grain component
in the absence of collisional fragmentation. In coagulation-fragmentation
equilibrium collisional fragmentation produces dust grains at the same rate as
they are swept up. But the sweep-up time-scale can also be associated with a
characteristic recycling time-scale. On the average dust fragments spend the
time $t_{\rm sweep}$ in the grain component, before they are incorporated into
a boulder. We have calculated the recycling time-scale for runs A and B, based
on \Eq{eq:trecyc}, in \Tab{t:results}. The grains have a recycling time-scale
of around three orbits in the two runs where dust fragments can not leave the
boulder layer. Thus the grains have been through approximately 100
agglomeration-destruction cycles during the course of runs A and B. We return
to the recycling time-scale in models where dust fragments are allowed to
diffuse out of the mid-plane layer in \S\ref{s:recyc2}.

\section{Diffusion}
\label{s:diffusion}

In this section we generalise the analytical model of \S\ref{s:analytics} to
1-D. Adding a $z$-direction to the problem and exposing the dust particles to
turbulent diffusion and vertical gravity yields the following equation system
for $\rho_1=m_1 n_1$ and $\rho_2=m_2 n_2$:
\begin{eqnarray}
  \frac{\dpa \rho_1}{\dpa t} &=&
      - \rho_1 \rho_2 \sigma_{12} v_{12}\frac{1}{m_2}
      + \rho_2^2 \sigma_{22} v_{22} \frac{1}{m_2} \nonumber \\
      && \hspace{2.0cm}- \frac{\dpa (w_1 \rho_1)}{\dpa z}
         + D_1 \frac{\dpa}{\dpa z}
           \left[\rho_{\rm g} \frac{\dpa (\rho_1/\rho_{\rm g})}{\dpa z} \right]
      \label{eq:drho1dt_diff} \, , \\
  \frac{\dpa \rho_2}{\dpa t} &=&
      + \rho_1 \rho_2 \sigma_{12} v_{12}\frac{1}{m_2}
      - \rho_2^2 \sigma_{22} v_{22} \frac{1}{m_2} \nonumber \\
      && \hspace{2.0cm}- \frac{\dpa (w_2 \rho_2)}{\dpa z}
         + D_2 \frac{\dpa}{\dpa z}
        \left[\rho_{\rm g} \frac{\dpa (\rho_2/\rho_{\rm g})}{\dpa z} \right] \,.
      \label{eq:drho2dt_diff}
\end{eqnarray}
Here $D_1$ and $D_2$ is the turbulent diffusion coefficient of grains and
boulders, respectively, and $w_1$ and $w_2$ are the vertical velocities,
assumed to be in equilibrium between gravity and drag force with
\begin{equation}\label{eq:wzsett}
  w_i = -\frac{z}{t_{\rm sett}} \approx -\frac{\varOmega_{\rm K}^2
  \tau_i}{1+\varOmega_{\rm K}^2\tau_i^2} z \, .
\end{equation}
This approximate expression recovers the terminal velocity of the grains in the
small friction time regime, $w_i=-\tau_i\varOmega_{\rm K}^2 z$, and the
settling time of oscillating particles in the large friction time regime,
$t_{\rm sett}=1/\tau_i$ \citep{YoudinLithwick2007}. The diffusion coefficient
$D_i$ depends on the friction time of the particles as
\citep[see][]{Carballido+etal2006,YoudinLithwick2007}
\begin{equation}\label{eq:Dt}
  D_i = \frac{D_0}{1+\varOmega_{\rm K}^2\tau_i^2} \, ,
\end{equation}
where $D_0$ is the diffusion coefficient of a passive scalar. We assume that
the turbulent mixing is independent of the height over the mid-plane, i.e.\
that $D_0$ is a constant. Ignoring the fragmentation and sweep-up terms of
\Eqs{eq:drho1dt_diff}{eq:drho2dt_diff} allows for a simple equilibrium between
diffusion and sedimentation,
\begin{equation}\label{eq:rhoi_gauss}
  \rho_i(z) = \frac{\varSigma_i}{\sqrt{2 \pi} H_i} \exp[-z^2/(2 H_i^2)] \, ,
\end{equation}
where the scale-height $H_i$ obeys the relation
\begin{equation}\label{eq:Hi}
  \frac{1}{H_i^2} = \frac{1}{H^2}
                  + \frac{\varOmega_{\rm K}^2 \tau_i}{D_0} \, , \\
\end{equation}
and $\varSigma_i$ is the total column density of solids of type $i$. Here the
denominators of \Eqs{eq:wzsett}{eq:Dt} have cancelled, making the short
friction time scale height expression (Eq.\ \ref{eq:Hi}) valid for all particle
sizes. However, the equilibrium solution in \Eqss{eq:rhoi_gauss}{eq:Hi} is
formally only valid when the friction time $\tau_i$ is assumed constant with
height over the mid-plane, an assumption that breaks down when considering the
dust distribution over several scale heights \citep{DullemondDominik2004}.
\begin{figure*}
  \begin{center}
    \includegraphics[width=0.3\linewidth]{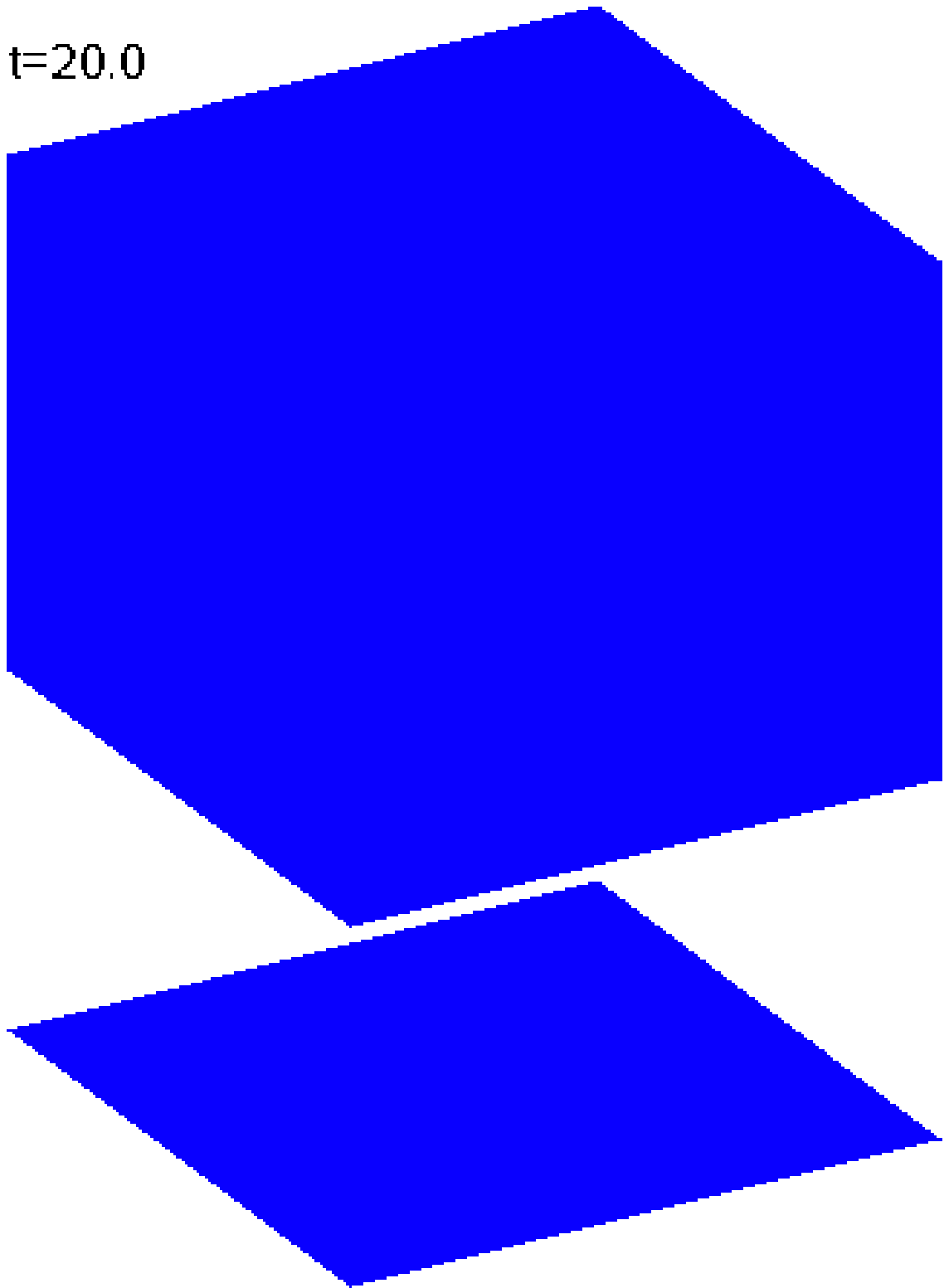}
    \includegraphics[width=0.3\linewidth]{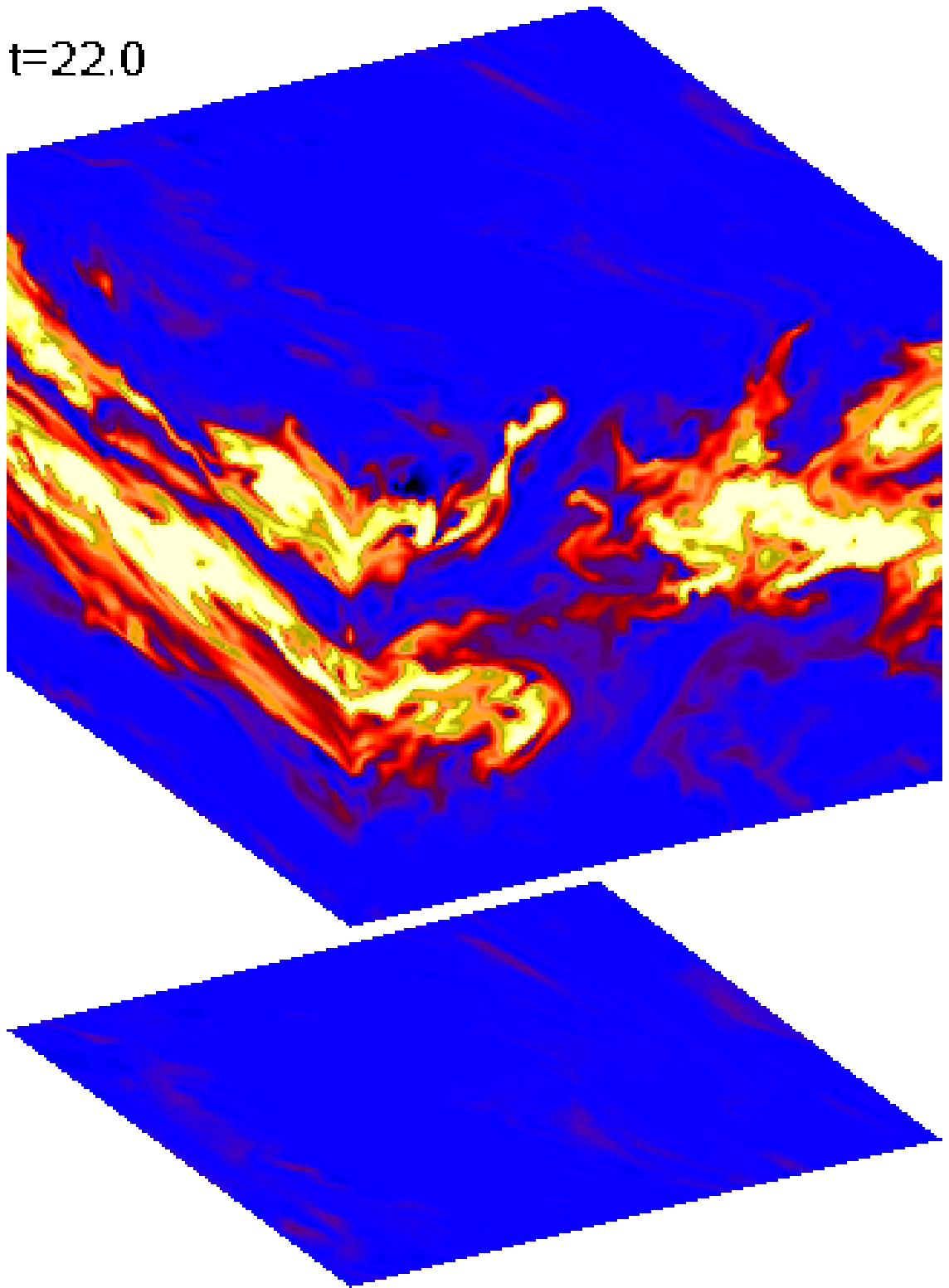}
    \includegraphics[width=0.3\linewidth]{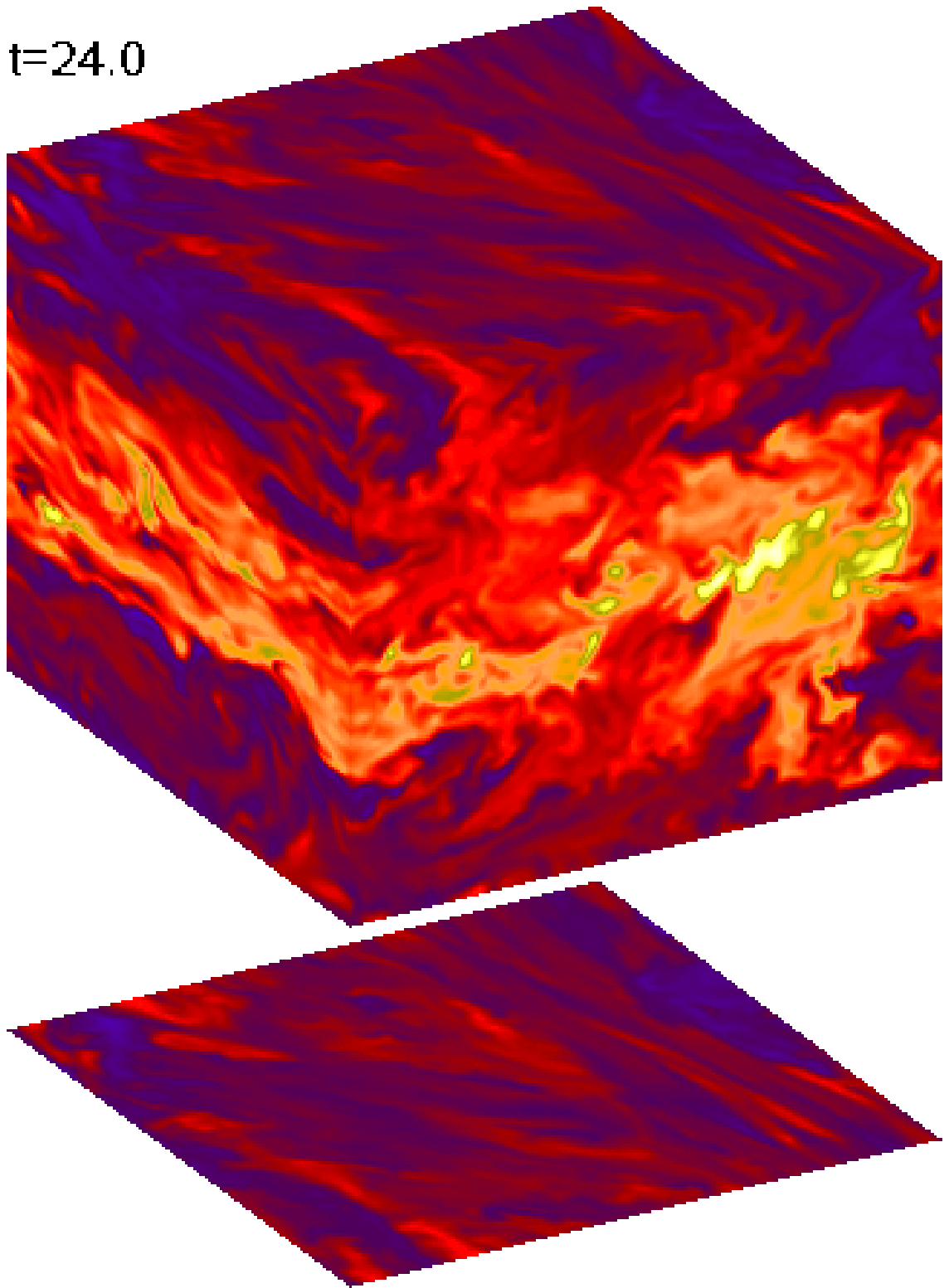}
    \includegraphics[width=0.3\linewidth]{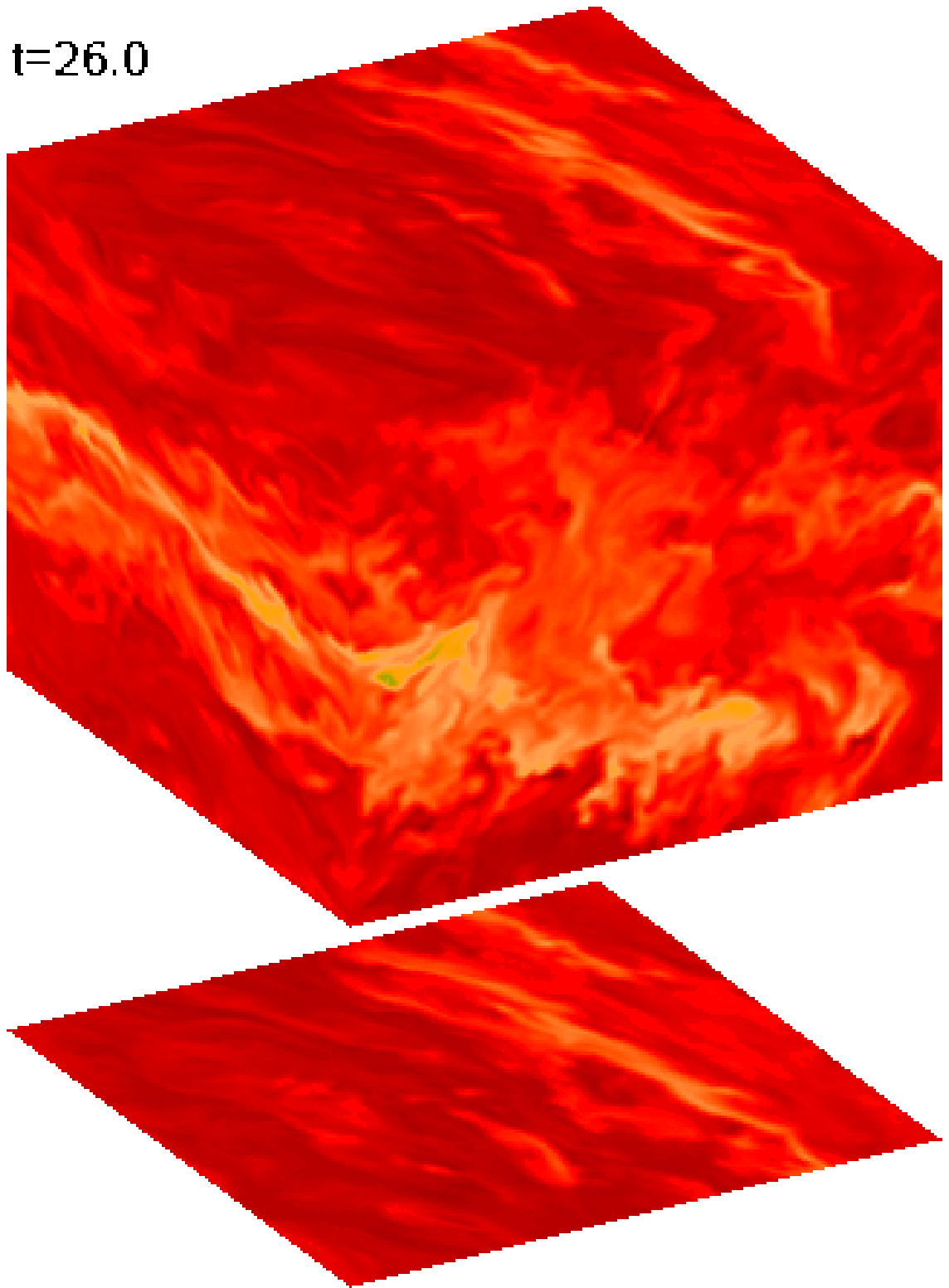}
    \includegraphics[width=0.3\linewidth]{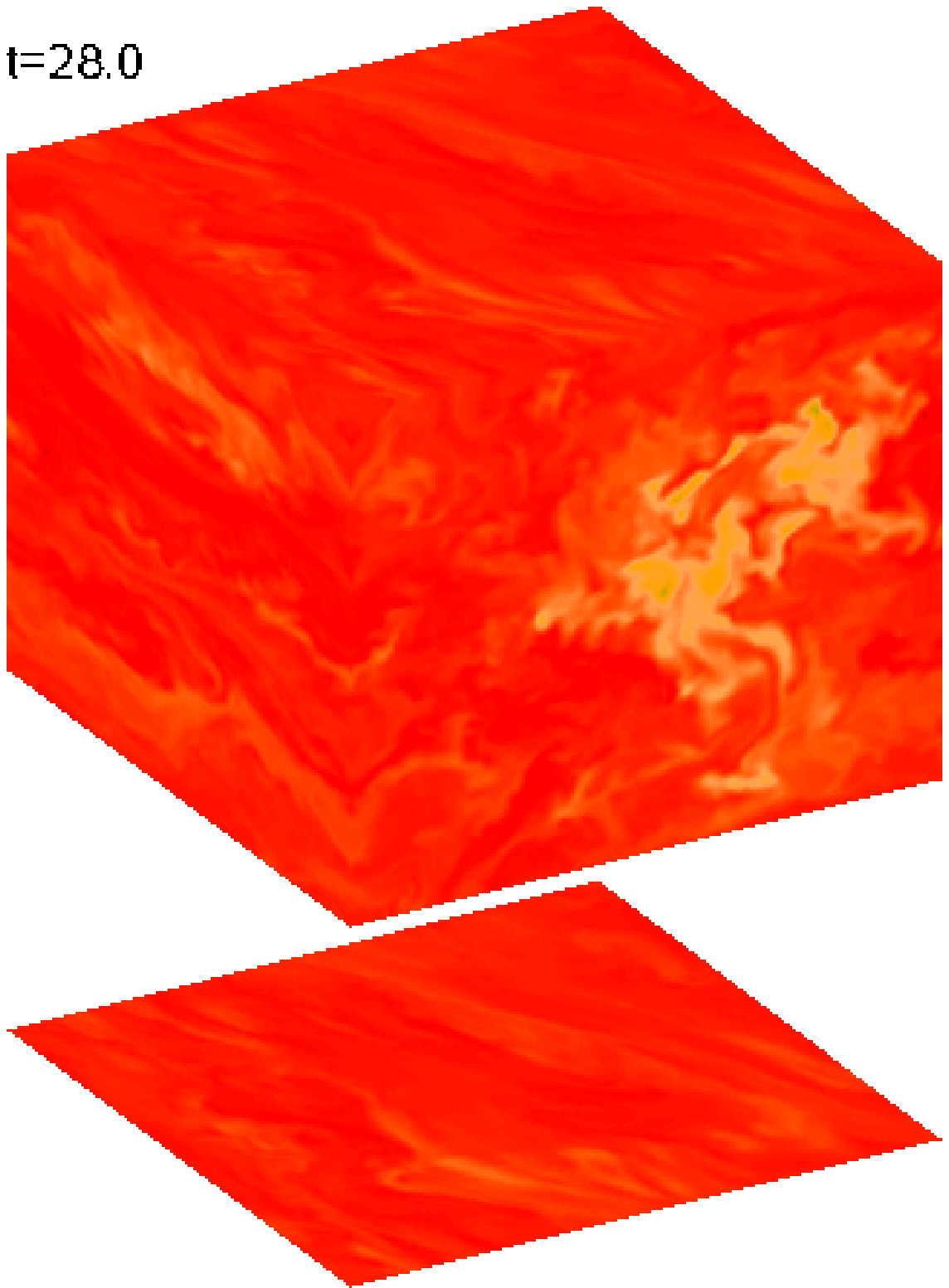}
    \includegraphics[width=0.3\linewidth]{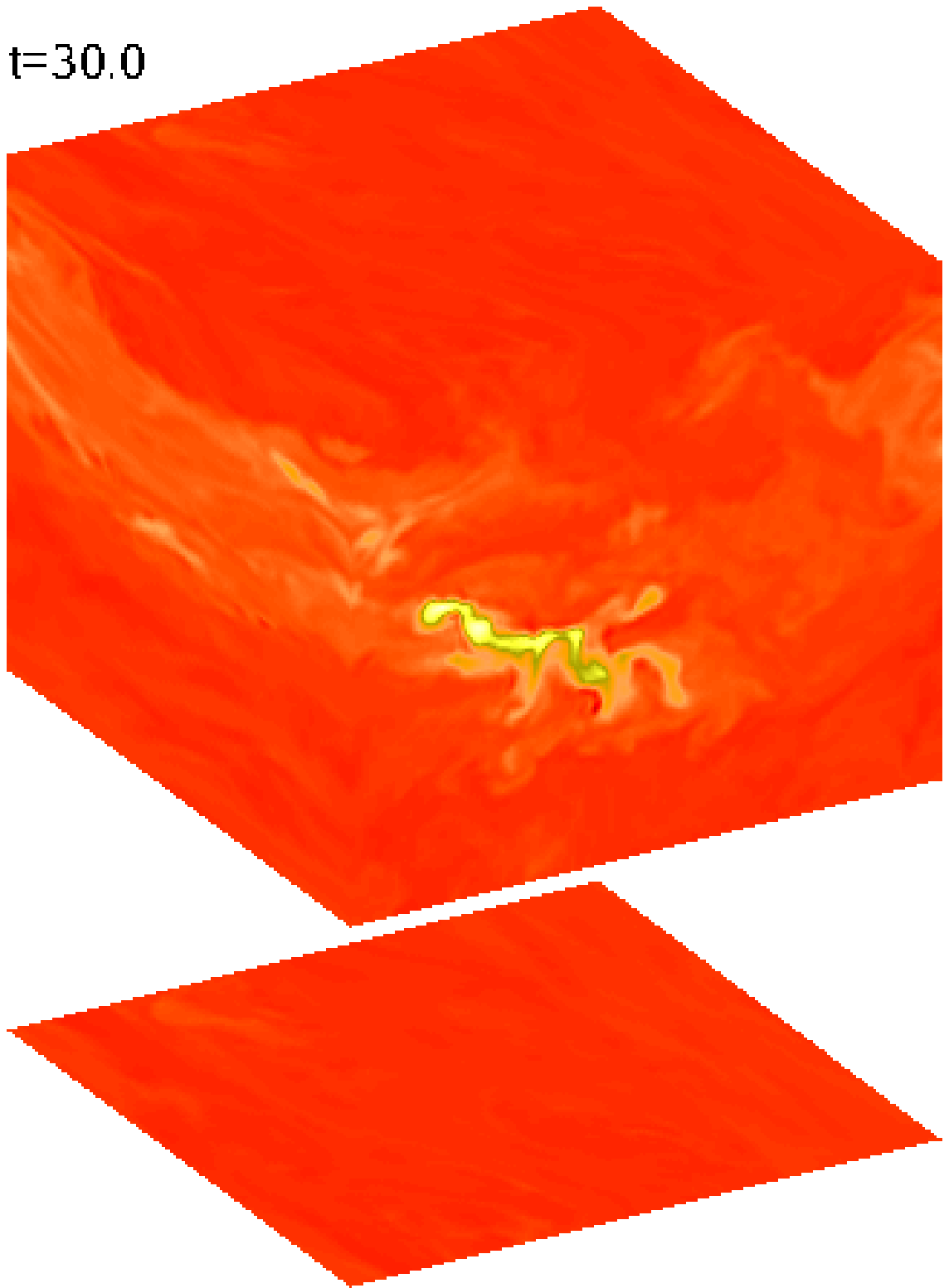}
  \end{center}
  \caption{Time series of the concentration of small dust grains for a run with
    $\alpha=0.01$ where the boulders lie in a thin layer around the mid-plane
    of the disc (run F). Fragments form when boulders collide in the mid-plane,
    but the produced dust grains escape from the mid-plane due to turbulent
    diffusion, spreading eventually evenly over the box.}
  \label{f:fragmentation_panels_gzp}
\end{figure*}

Integrating \Eqs{eq:drho1dt_diff}{eq:drho2dt_diff} over the entire $z$-space
yields dynamical equations for the column densities $\varSigma_1$ and
$\varSigma_2$ instead,
\begin{eqnarray}
  \frac{\dpa \varSigma_1}{\dpa t} &=&
      \int_{-\infty}^\infty \left(
      -\rho_1 \rho_2 \sigma_{12} v_{12}\frac{1}{m_2}
      + \rho_2^2 \sigma_{22} v_{22} \frac{1}{m_2} \right) \de z
      \label{eq:dsigma1dt} \, , \\
  \frac{\dpa \varSigma_2}{\dpa t} &=&
      \int_{-\infty}^\infty \left(
      +\rho_1 \rho_2 \sigma_{12} v_{12}\frac{1}{m_2}
      - \rho_2^2 \sigma_{22} v_{22} \frac{1}{m_2} \right) \de z
      \label{eq:dsigma2dt} \, ,
\end{eqnarray}
where the derivative terms have vanished because the column density of solids
can not be changed by sedimentation and vertical diffusion. Inserting the
sedimentation-diffusion equilibrium solution from \Eq{eq:rhoi_gauss} into
\Eqs{eq:dsigma1dt}{eq:dsigma2dt} and searching for $\varSigma_1/\varSigma_2$
that gives $\dot{\varSigma}_1=\dot{\varSigma}_2=0$ yields
\begin{equation}\label{eq:Sigmarat_equi}
  \frac{\varSigma_1}{\varSigma_2}
      = \frac{\sigma_{22}}{\sigma_{12}} \frac{v_{22}}{v_{12}}
        \frac{\sqrt{1+H_1^2/H_2^2}}{\sqrt{2}}
\end{equation}
as an extension to \Eq{eq:rhorat_equi}. If $\tau_1=\tau_2$, then $H_1=H_2$
according to \Eq{eq:Hi}, and
$\varSigma_1/\varSigma_2=(\sigma_{22}/\sigma_{12})(v_{22}/ v_{12})$, completely
equivalent to the 0-D case. Combining Eqs.\ (\ref{eq:rhoi_gauss}),
(\ref{eq:Hi}) and (\ref{eq:Sigmarat_equi}) and inserting $\rho_1(z=0)$ in
\Eq{eq:da2dt} yields the radius growth of boulders in the mid-plane as
\begin{equation}\label{eq:a2dot_diff}
  \dot{a}_2 = \frac{\varSigma_{1+2}/(\sqrt{2\pi}H_1)}{\rho_\bullet}
      \frac{v_{22}}{[2/(1+H_1^2/H_2^2)]^{1/2}+4v_{22}/v_{12}}
      \, ,
\end{equation}
the 1-D generalisation of \Eq{eq:da2dt_equi}. The growth rate of the boulders
is more or less inversely proportional to $H_1$ (note that the $H_1^2/H_2^2$
term in the denominator of Eq.\ [\ref{eq:a2dot_diff}] has only little influence
on the growth rate for $H_1\gg H_2$). Thus going from $H_1=H_2$ to, say,
$H_1=10 H_2$, a reasonable value for small fragments, decreases the growth rate
by coagulation-fragmentation by a dramatic order of magnitude.
\begin{figure*}
  \begin{center}
    \includegraphics[width=\linewidth]{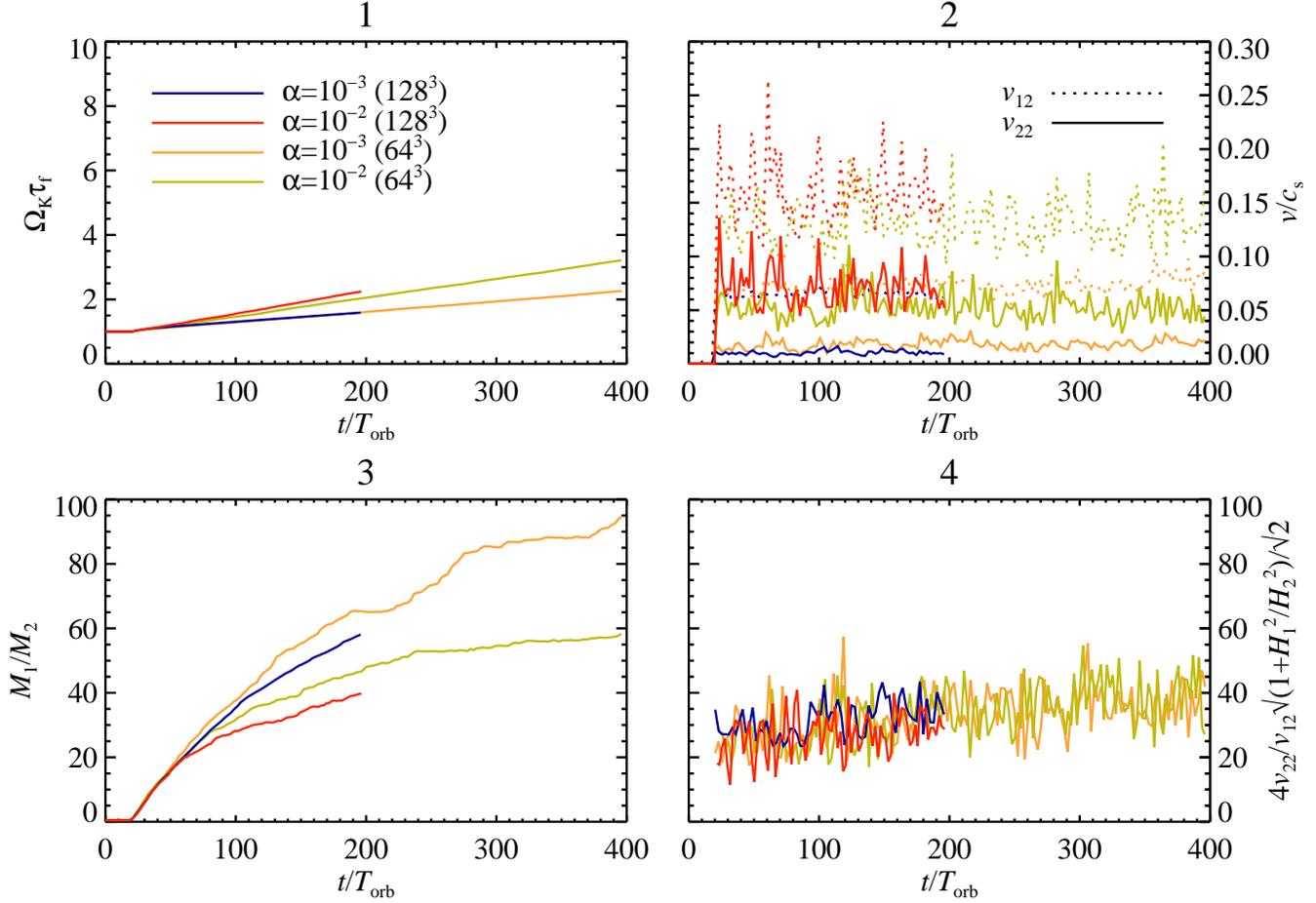}
  \end{center}
  \caption{Growth and fragmentation of boulders in models where the dust
    fragments are allowed to diffuse out of the boulder layer. The radius of
    the boulders increases much slower than in \Fig{f:dvp_t}, since fragments
    created in catastrophic collisions easily escape from the mid-plane layer
    and spread over the entire vertical extent of the disc. The ratio of
    fragments to boulders (panel 3) approximately matches the value expected
    from analytical derivations (panel 4) based on the different equilibrium
    scale height of boulders and fragments. Due to the slow growth boulders
    will have drifted into the inner disc to evaporate there before the meter
    barrier would eventually be crossed after a few thousand orbits.}
  \label{f:dvp_t_gzp}
\end{figure*}

The validity of the Gaussian solution (Eq.\ [\ref{eq:rhoi_gauss}]) can be
quantified by comparing the relevant time-scales of
\Eqs{eq:drho1dt_diff}{eq:drho2dt_diff} -- the time-scale for fragments to
diffuse out of the mid-plane layer, $t_{\rm diff} = H_2^2/D_{\rm t}$, and the
collisional time-scale, $t_{\rm coll} = m_2/(\rho_2 \sigma_{22} v_{22})$. The
Gaussian solution is valid when $t_{\rm diff}\ll t_{\rm coll}$, giving
\begin{equation}
  \varOmega_{\rm K}^2 \tau_2^2 \gg
      \frac{\rho_2}{\rho_{\rm g}} \frac{v_{22}}{c_{\rm s}} \, ,
\end{equation}
where we used $\tau_2=(a_2 \rho_\bullet)/(c_{\rm s} \rho_{\rm g})$ to translate
the sedimentation time-scale into an Epstein friction time. Using further the
sedimentation-diffusion equilibrium expression for the mid-plane density
$\rho_2/\rho_{\rm g}=\epsilon_0\sqrt{\varOmega_{\rm K}\tau_{\rm f}/\delta}$,
where $\epsilon_0$ is the global solids-to-gas ratio of the boulder component
\citep[see e.g.][]{JohansenKlahrHenning2006}, yields
\begin{equation}
  \varOmega_{\rm K} \tau_{\rm f} \gg \delta^{-1/3} \left( \epsilon_0
  \frac{v_{22}}{c_{\rm s}} \right)^{2/3}
\end{equation}
with fully independent parameters (both the collision speed $v_{22}$ and the
diffusion coefficient $\delta$ of course depend on the strength of the
turbulence, so these two must be chosen consistently). For $\delta =10^{-3}$,
$\epsilon_0=0.01$ and $v_{22}/c_{\rm s}=0.02$ the criterion for the validity of
the Gaussian solution is $\varOmega_{\rm K} \tau_2 \gg 0.03$, in accordance
with our modelling of component 2 as boulders with $\varOmega_{\rm K} \tau_{\rm
f} \gtrsim 1$.

The simple model presented in \S\ref{s:clumping} for the effect of clumping on
coagulation-fragmentation also gives a new perspective on the effect of
sedimentation [\Eqs{eq:Sigmarat_equi}{eq:a2dot_diff}]. Decreasing the scale
height of boulders from $H_2=H_1$ to $H_2\ll H_1$ leads to a steep rise in
$\varSigma_1/\varSigma_2$ [\Eq{eq:Sigmarat_equi}], but only to a mild increase
in radius growth [\Eq{eq:a2dot_diff}]. Thus any non-homogeneity of the boulder
layer (be it due to concentrations in transient gas high pressure or due to
sedimentation) has little effect on the growth rate of the boulders, but may
reduce the bulk density of boulders drastically. Taking instead the scale
height of fragments $H_1$ and increasing it from $H_1=H_2$ to $H_1\gg H_2$, a
transition which is similar to the one occurring from models A-B to models C-F,
leads to a sharp decrease in both growth rate and column density of the boulder
component.

\section{Results including leaking}
\label{s:sed}

Having found in the preceding section an approximate analytical solution for
the coagulation-fragmentation equilibrium in the case where dust fragments are
free to leave the sedimentary boulder layer, we return now to the results of
numerical simulations. In runs C-F we set the solids-to-gas ratio of boulders
and fragments to the more canonical value $0.01$ and expose the boulders to
vertical gravity. We give the boulders time to settle to the mid-plane from
$t=10 T_{\rm orb}$ to $t=20 T_{\rm orb}$, so that an equilibrium sedimentary
layer has formed when sweep-up and collisional fragmentation are turned on at
$t=20 T_{\rm orb}$.

In \Figs{f:fragmentation_panels_gzp}{f:dvp_t_gzp} we show the time evolution of
fragments and boulders in models where the boulders have sedimented out of the
gas to establish a thin layer around the mid-plane of the box. Fragments form
in catastrophic collision between boulders in the mid-plane, but are quickly
carried away from the mid-plane layer by the turbulent gas and are eventually
well-mixed throughout the box (\Fig{f:fragmentation_panels_gzp}).
\Fig{f:dvp_t_gzp} shows, for two different numerical resolutions, that the
boulders grow a factor ten slower than in the case where dust fragments were
not allowed to leave the boulder layer (\Fig{f:dvp_t}). Also the mass of
fragments is huge because the catastrophic collisions have to keep up with the
turbulent transport of fragments away from the mid-plane. The fourth panel of
\Fig{f:dvp_t_gzp} shows the analytical expectation value for $M_1/M_2$,
following \Eq{eq:Sigmarat_equi}, by setting the expected scale height of
fragments, $H_1$, equal to the gas scale height and using the collision speeds
from the second panel. There is a bit more mass in fragments than expected,
which may, as in the case where dust fragments were not allowed to leave the
boulder layer shown in \Fig{f:dvp_t}, be explained by the clumpy structure of
the boulder layer (see \S\ref{s:clumping}).

The turbulent transport reduces the column density of boulders to approximately
1-2\% of the total column density of the solids. The growth rate of the Stokes
number of the boulders is around $0.003\ldots0.007$ per orbit, with the higher
values appearing in the strongly turbulent $\alpha=10^{-2}$ runs. Under all
circumstances this is way too low to compete with radial drift which occurs on
a time-scale of a few ten orbits. Even if the radial drift time-scale is
ignored it would take around 1000 orbits to grow to ${\rm St}=10$. The
extremely small column density of the boulders corresponds to around 1 boulder
per (10 km)$^2$. It is somewhat surprising that so few boulders can populate
the entire vertical extent of the disc with small dust grains.
\cite{Eisner+etal2006} recently modelled submicron-sized dust grains in the
inner part of the transition disc TW Hya and found that the lifetimes of these
grains against radiation pressure is much shorter than the age of the system.
One can speculate that the source of these small grains could be boulders
(drifting through the inner disc or permanently situated there) creating
observable amounts of fragments as they collide.
\begin{figure}[!t]
  \begin{center}
    \includegraphics[width=\linewidth]{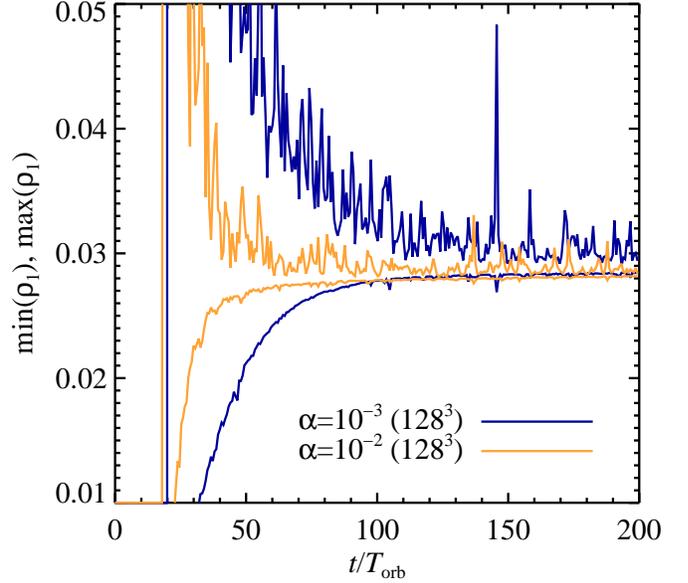}
  \end{center}
  \caption{The minimum and maximum bulk density of fragments, $\rho_1$, in the
    box as a function of time for runs E and F. Collisional fragmentation is
    turned on at $t=20 T_{\rm orb}$, followed by an initial peak in the maximum
    density of fragments. Turbulent mixing nevertheless leads to an almost
    homogeneous state with only very little difference between minimum and
    maximum concentrations of fragments. The case $\alpha=10^{-2}$ has less
    difference between minimum and maximum concentrations than for
    $\alpha=10^{-3}$ due to stronger mixing in the first case.}
  \label{f:rho1minmax_t}
\end{figure}

The minimum and maximum bulk density of fragments in the gas is shown in
\Fig{f:rho1minmax_t}. After collisional fragmentation is turned on at $t=20
T_{\rm orb}$, there is a sharp increase in the maximum concentration of
fragments, but turbulent mixing eventually leads to a state where the minimum
and maximum concentrations are within 3\% of each other for $\alpha=10^{-2}$
and within 6\% of each other for $\alpha=10^{-3}$. It may be surprising that
the difference is so little, but it is again because the collisional time-scale
is much longer than the diffusion time-scale. That also explains why the
$\alpha=10^{-2}$ has less difference between minimum and maximum concentrations
than $\alpha=10^{-3}$ -- the stronger mixing in the highly turbulent case evens
out concentration differences more effectively. We note that although the
average solids-to-gas ratio in the disc is assumed to be $\epsilon_0=0.02$, the
fragments reach an average bulk density of $\epsilon_0=0.029$. This is an
artificial effect of the limited vertical extent of the box: boulders have
sedimented out of the regions outside of the box, but fragments are not allowed
to escape back out the box, hence the average density of fragments is kept
artificially high.

\subsection{Recycling time-scale revisited}
\label{s:recyc2}

We introduced the recycling time-scale in \S\ref{s:recyc}. With diffusion of
dust fragments from the boulder layer suppressed, the dust grains would be free
for only around three orbits before being incorporated into a boulder. We now
derive a similar expression for the recycling time-scale, valid in the case
where dust fragments are allowed to diffuse out of the thin mid-plane layer
where they originate in collisions between boulders.

Assuming that the density dependence on height over the mid-plane is Gaussian
for both species (Eq.\ [\ref{eq:rhoi_gauss}]) we can evaluate the integral in
\Eq{eq:dsigma1dt} analytically and get the recycling time-scale
\begin{equation}
  t_{\rm recyc} = \frac{\sqrt{2\pi}\sqrt{H_1^2+H_2^2}}{\varSigma_{2}}
  \frac{m_2}{\sigma_{12}v_{12}} \, .
\end{equation}
Notice that the recycling time-scale for $H_1=H_2$ is $\sqrt{2}$ times longer
than in the case where dust fragments and boulders were not allowed to separate
vertically, because, although the densities in the very mid-plane are exactly
as in the 0-D case, the vertically averaged collision rate is smaller, which
leads to a somewhat longer recycling time-scale.

Inserting the coagulation-fragmentation equilibrium from \Eq{eq:Sigmarat_equi}
yields the recycling time-scale in units of the friction time as
\begin{eqnarray}
  \frac{t_{\rm recyc}}{\tau_{\rm f}} &=&
    \frac{1+4(v_{22}/v_{12})
    \left(\sqrt{1+H_1^2/H_2^2}/\sqrt{2}\right)}{3} \times \nonumber \\
    && \hspace{1.0cm}
    \left[ \frac{v_{12}}{c_{\rm s}} \right]^{-1}
    \left[
    \frac{\varSigma_{1+2}/\left(\sqrt{2\pi}\sqrt{H_1^2+H_2^2}\right)}{\rho_{\rm
    g}} \right]^{-1} \label{eq:trecycsed} \, ,
\end{eqnarray}
completely equivalent to \Eq{eq:trecyc}. Using
$\varSigma_{1+2}=0.02\varSigma_{\rm g}=0.02\sqrt{2\pi}$ in units where the
mid-plane gas density is unity, together with $H_1/H_{\rm g}=1$ and the
diffusion-sedimentation equilibrium $H_2/H_{\rm g}=\sqrt{\delta
/(\varOmega_{\rm K}\tau_{\rm f}})$, we have calculated the recycling time-scale
based on \Eq{eq:trecycsed} in \Tab{t:results}. Allowing dust fragments to leak
out of the mid-plane layer leads to a dramatic increase by two orders of
magnitude in the recycling time-scale, which is now 200-300 orbits for the
$\alpha=10^{-2}$ models and approaching a thousand orbits for the
$\alpha=10^{-3}$ runs. Not only does the vertical diffusion decrease the amount
of boulders in the mid-plane by approximately an order of magnitude, leading to
a much longer sweep-up time-scale, but the total amount of dust grains is also
higher, so that it takes much longer time for the average dust grain to
encounter a boulder.

\section{Summary and discussion}
\label{s:conclusions}

We have proposed a simple two-component model for the growth and collisional
destruction of boulders in protoplanetary discs. Fragments produced in
catastrophic collisions between boulders are swept up by other boulders,
leading to a continuous growth towards larger bodies. An analytical equilibrium
solution to the dynamical equations predicts that the boulder radius can grow
as quickly as a few mm per year. The promise of this method to provide growth
rates that can compete with the radial drift is nevertheless compromised by the
inclusion of turbulent transport of fragments out of the boulder layer. The
sedimentary mid-plane layer loses the produced fragments to the boulder-free
parts of the disc, eventually grinding down the boulder component to
insignificant masses and reducing the growth rate drastically. Thus what
initially seemed to have the potential to provide an efficient growth phase of
boulders turned into something like a worst-case scenario with the inclusion of
turbulent transport.

One can think of ways by which boulders may still penetrate the meter barrier
by sweep-up of small fragments:
\begin{itemize}
  \item Fragments are larger
  \item Boulder collisions do not lead to destruction
  \item Protoplanetary discs are less turbulent than assumed
  \item Turbulence is confined to the mid-plane
  \item Radial drift is weaker in nature than thought
\end{itemize}

If fragments are large enough that they do not couple instantaneously (compared
to an orbital time-scale) to the gas, then the turbulent transport away from
the sedimentary mid-plane layer is slowed down. Having such large fragments
would in principle put coagulation-fragmentation growth back on track. BDH
presented models with a much more advanced collisional fragmentation model,
where the results of catastrophic collisions are distributed with a power law,
but this still did not lead to a break-through of the meter barrier. If, on the
other hand, boulders do not fragment at all, but merely bounce off each other,
then the whole concept of coagulation-fragmentation growth breaks down, and the
stage is left to coagulation of equal-sized boulders and/or self-gravity in the
boulder component \citep{Johansen+etal2007}.

Turbulence plays an interesting double role in the coagulation-fragmentation
process. The relative speed between the boulders, which leads to their
destruction and the continuous replenishment of the grains, is induced by the
marginal coupling of the boulders to the turbulent gas motion. On the other
hand turbulent diffusion drains the mid-plane layer of fragments and reduces
the growth rate of the boulders. This is in some opposition to the role that is
normally attributed to turbulence in the coagulation process. Here a dense
mid-plane layer and reduced collisional fragmentation are desired, and
turbulence counteracts both. Coagulation-fragmentation growth, on the other
hand, benefits directly from stronger turbulence and higher collision speeds
\citep[although the sweep-up efficiency can be put into question when the
relative speed between dust grains and boulders increases beyond a few ten
meters per second, see discussions in][]{Wurm+etal2001,Wurm+etal2005}, while it
is indifferent to sedimentation and clumping of the boulder layer.

The presented simulations all have a simplified space-filling turbulence of
magnetic origin. If turbulence would instead be confined to the mid-plane of
the disc, as assumed in \cite{Weidenschilling1997}, then the transport of
grains away from the mid-plane can be reduced, while at the same time the
collision speeds can be kept high. The Kelvin-Helmholtz and streaming
instabilities associated with the sedimentation of solids are nevertheless not
Keplerian shear instabilities \citep{YoudinGoodman2005,YoudinJohansen2007} and
thus cannot explain the observed accretion rates of young stars. If, on the
other hand, there is a region around the mid-plane where the magnetic field
does not couple well enough to the gas to have magnetorotational instability
\citep{Gammie1996,Oishi+etal2007}, then one could have lower turbulence in the
mid-plane and much less loss of fragments to the boulder-free parts of the
disc, but as discussed above, a net decrease of turbulence in the mid-plane has
a negative impact on the coagulation-fragmentation growth. Dust grains are a
major catalyst for recombination of ionised species \citep{Sano+etal2000}. This
leads to an interesting coupling between dust and turbulence whereby dusty
regions would have weaker turbulent motion, and thus less diffusion, than
dust-free regions. We plan to include such effects in a future model.

Collisional fragmentation is actually not the real problem for planetesimal
formation, given that fragments are readily incorporated in the few lucky
boulders that avoid catastrophic collisions. The problem arises because the
time-scale of growth by coagulation-fragmentation is so long that all material
will have been flushed through the disc long before being able to grow big
enough to decouple from the gas. Radial drift in the minimum mass solar nebula
reaches 5\% of the local sound speed \citep{Weidenschilling1977}, yielding a
drift time-scale of a few ten orbits, much too short for any significant radius
growth.

There is a mounting observational evidence that radial drift may not be as big
in actual discs as predicted from theoretical arguments. \cite{Rettig+etal2006}
measured the abundance of gas and dust in discs of millions of years in age and
were able to explain the inclination dependence of the dust-to-gas ratio from a
combination of grain growth and sedimentation, still within the framework of
the minimum mass solar nebula with a solids-to-gas ratio of 0.01. In the models
of drift and coagulation presented in BDH, on the other hand, the outer disc is
cleared of dust in a few hundred thousand years. The observed presence of
cm-sized solids in the outer parts of protoplanetary discs
\citep{Wilner+etal2000,Testi+etal2003,Rodmann+etal2006,Lommen+etal2007} is at
best marginally consistent with theoretical life-times of such grains due to
radial drift \citep{Brauer+etal2007}. Maybe the most promising way to stop
radial drift is to have radial pressure bumps in the disc \citep{Whipple1972}.
These bumps can arise from first principles in 3-D simulations of the dynamics
of protoplanetary discs, e.g.\ spiral arms of self-gravitating discs
\citep{Rice+etal2004} or long-lived high pressure regions in magnetorotational
turbulence \citep{FromangNelson2005,JohansenKlahrHenning2006}. Any pressure
bump must compete with the global pressure gradient and produce a net zero or
positive gradient. This requirement is nevertheless not necessarily very
difficult to fulfil. The gas pressure typically falls 10\% over one scale
height in the radial direction, a value with which even subsonic turbulence can
easily compete. The challenge is to have long-lived pressure enhancements, an
issue which is still debated for the case of magnetorotational turbulence
\citep{FromangNelson2006}. Radial drift may also be significantly reduced in a
dense boulder-dominated mid-plane layer where the gaseous headwind is reduced
as the gas is dragged along with the boulders \citep{Nakagawa+etal1986}.

The local absence of radial drift would not only solve the time-scale problem,
but also reduce differential radial drift, which was shown by BDH to lead to
destruction of the boulders due to collisions at speeds between 10 and 50 m/s
with slightly smaller boulders. The process of crossing from 1 m to 10 m would
still be very inefficient, since the turbulent transport would cause 99\% of
the solid mass to be present in small fragments, but given that planet
formation in our solar system took place over millions of years \citep[see
review by][]{TrieloffPalme2006}, inefficient planetesimal formation may
actually be desired to comply with meteoritic evidence. Once a few
extraordinarily lucky bodies would cross the meter barrier, they could grow big
enough to even sweep up the boulders still lying around in the mid-plane and
thus continue to grow towards full-fledged protoplanets and later gas giant
cores and terrestrial planets.

\begin{acknowledgements}
  Part of this work was supported by the EU planets network. We are grateful to
  J\"urgen Blum and Andrew Youdin for inspiring discussions on
  coagulation-fragmentation growth. We would like to thank the referee, Dr.\
  Stuart Weidenschilling, for a thorough reading of the manuscript and for many
  suggestions of improvements.
\end{acknowledgements}

\begin{appendix}

\section{Stability analysis}\label{s:stability}

In this appendix we consider the stability of the equilibrium solution to
\Eq{eq:dn1dt}. We find that the amplitude of any (arbitrarily large)
perturbation to the equilibrium will decrease exponentially with time, so that
\Eq{eq:rhorat_equi} is a (both linearly and non-linearly) stable solution to
the coagulation-fragmentation problem.

We start by rewriting \Eq{eq:dn1dt} slightly by multiplying with $m_1$ and
inserting $\rho_{1+2}=\rho_1+\rho_2$ (mass conservation) to avoid any reference
to $\rho_2$. The resulting dynamical equation is
\begin{equation}\label{eq:drho1dt}
  \frac{\dpa \rho_1}{\dpa t} = -\rho_1 \frac{\rho_{1+2}-\rho_1}{m_2}
  \sigma_{12} v_{12} + \frac{(\rho_{1+2}-\rho_1)^2}{m_2} \sigma_{22}
  v_{22} \, .
\end{equation}
We linearise this equation by writing the density of fragments $\rho_1$ as
\begin{equation}\label{eq:linexp}
  \rho_1 = \overline{\rho}_1 + \rho_1'(t) \, ,
\end{equation}
where $\overline{\rho}_1$ is the equilibrium solution to \Eq{eq:drho1dt},
\begin{equation}\label{eq:rho1_equi}
  \overline{\rho}_1 = \rho_{1+2} \frac{\sigma_{22} v_{22}}
  {\sigma_{12} v_{12} + \sigma_{22} v_{22}} \, ,
\end{equation}
and $\rho_1'$ is an infinitesimal perturbation to this equilibrium density. We
ignored the other equilibrium solution, $\overline{\rho}_1=\rho_{1+2}$, because
that state is not accompanied by any radius increase of the boulders (see
discussion below). The expression in \Eq{eq:rho1_equi} is completely similar to
\Eq{eq:rhorat_equi}, but we avoided approximating $\sigma_{22} \approx 4
\sigma_{12}$ for generality reasons. Inserting \Eq{eq:linexp} into
\Eq{eq:drho1dt} yields the dynamical equation
\begin{eqnarray}
  \frac{\dpa \rho_1'}{\dpa t} &=&
    -(\overline{\rho}_1+\rho_1')
    \frac{\rho_{1+2}-\overline{\rho}_1-\rho_1'}{m_2}
    \sigma_{12} v_{12} \nonumber \\
    && \hspace{2.5cm} +
    \frac{(\rho_{1+2}-\overline{\rho}_1-\rho_1')^2}{m_2} \sigma_{22}
    v_{22}
\end{eqnarray}
for the density perturbation $\rho_1'$. Using the fact that $\overline{\rho}_1$
in itself satisfies \Eq{eq:drho1dt} and ignoring terms of second order in the
perturbed density yields the linearised dynamical equation
%\begin{equation}
%  \frac{\dpa \rho_1'}{\dpa t} =
%    -\frac{\sigma_{12} v_{12}}{m_2}
%    ( \overline{\rho}_1 \rho_{1+2} - \overline{\rho}_1 \overline{\rho}_1 -
%      \overline{\rho}_1 \rho_1' + \rho_1' \rho_{1+2} -
%      \rho_1' \overline{\rho}_1 -\rho_1' \rho_1' ) +
%    \frac{\sigma_{22} v_{22}}{m_2}
%    ( \rho_{1+2} \rho_{1+2} - \rho_{1+2} \overline{\rho}_1 -
%      \rho_{1+2} \rho_1' - \overline{\rho}_1 \rho_{1+2} +
%      \overline{\rho}_1 \overline{\rho}_1 + \overline{\rho}_1 \rho_1' -
%      \rho_1' \rho_{1+2} + \rho_1' \overline{\rho}_1 + \rho_1' \rho_1' )
%\end{equation}
\begin{eqnarray}
  \frac{\dpa \rho_1'}{\dpa t} &=&
  - \frac{\sigma_{12} v_{12}}{m_2}
  (-2 \overline{\rho}_1 \rho_1' + \rho_1' \rho_{1+2} ) \nonumber \\
  && \hspace{2.5cm} + \frac{\sigma_{22} v_{22}}{m_2}
  (-2 \rho_{1+2} \rho_1' + 2 \overline{\rho}_1 \rho_1') \, .
\end{eqnarray}
Inserting now the equilibrium solution $\overline{\rho}_1$ from
\Eq{eq:rho1_equi} gives, after some trivial algebraic manipulation, the final
equation for the evolution of the perturbed density as
\begin{figure}
  \begin{center}
    \includegraphics[width=\linewidth]{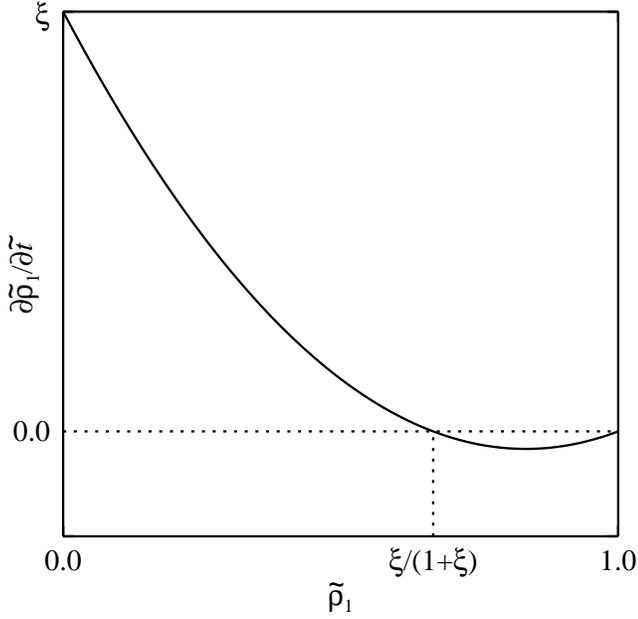}
  \end{center}
  \caption{The dependence of the time derivative of the normalised density of
    fragments, $\dpa \tilde{\rho}_1/\dpa \tilde{t}$, on the density,
    $\tilde{\rho}_1 \equiv \rho_1/\rho_{1+2}$, itself, following
    \Eq{eq:drho1tilde_dt}. All states tend towards the equilibrium state at
    $\tilde{\rho}_1=\xi/(1+\xi)$. Here $\xi \equiv \sigma_{22}
    v_{22}/(\sigma_{12} v_{12})$ is a parameter that depends on the
    collision speeds and on the cross section of the boulders. The equilibrium
    at $\tilde{\rho}_1=1$ is clearly unstable.}
  \label{f:drho1tilde_rho1}
\end{figure}
%\begin{equation}
%  \frac{\dpa \rho_1'}{\dpa t} = \frac{\rho_{1+2}}{m_2}
%  \left[ \sigma_{12} v_{12} \left( \frac{2 \sigma_{22}
%  v_{22}}{\sigma_{12} v_{12}+\sigma_{22} v_{22}} -1 \right)
%  + \sigma_{22} v_{22}
%  \left(-2 + \frac{2 \sigma_{22} v_{22}}{\sigma_{12} v_{12}+\sigma_{22}
%  v_{22}} \right) \right] \rho_1'
%\end{equation}
\begin{equation}
  \frac{\dpa \rho_1'}{\dpa t} =
    - \frac{\rho_{1+2}}{m_2} \sigma_{12} v_{12} \rho_1' \, .
\end{equation}
Thus any small variation from the equilibrium expression will decay
exponentially [following $\rho_1'(t) \propto \exp(-\omega t)$] at the rate
\begin{equation}
  \omega = \frac{\rho_{1+2} \sigma_{12} v_{12}}{m_2} \, .
\end{equation}
The corresponding decay time $m_2/(\rho_{1+2} \sigma_{12} v_{12})$ is,
not surprisingly, similar to the time-scale of the sweep-up process, but with
the the density of the fragments replaced by the total density $\rho_{1+2}$.

One may suspect that the equilibrium solution is also non-linearly stable, due
to the simple nature of the \Eq{eq:drho1dt}. Rewriting \Eq{eq:drho1dt} in terms
of the normalised density $\tilde{\rho}_1=\rho_1/\rho_{1+2}$, normalised time
$\tilde{t}=t/[m_2/(\rho_{1+2} \sigma_{12} v_{12})]$ and collision
parameter $\xi=\sigma_{22} v_{22}/(\sigma_{12} v_{12})$ yields
the evolution equation
\begin{equation}\label{eq:drho1tilde_dt}
  \frac{\dpa \tilde{\rho}_1}{\dpa \tilde{t}} = (1+\xi) \tilde{\rho}_1^2 +
    (-1-2\xi)\tilde{\rho}_1 + \xi \, .
\end{equation}
The second order polynomial $\dpa \tilde{\rho}_1/\dpa \tilde{t}=0$ has the
solutions
\begin{eqnarray}
  r_1 &=& 1 \, , \label{eq:r1} \\
  r_2 &=& \frac{\xi}{1+\xi}\, . \label{eq:r2}
\end{eqnarray}
In the first solution all the solid mass is bound in fragments -- the total
absence of boulders for this case means that coagulation-fragmentation growth
is disabled. The second solution is the same as in \Eq{eq:rho1_equi}. Since we
know that the polynomial in \Eq{eq:drho1tilde_dt} opens upwards and that there
is only one crossing of zero in the interval $[0,1[$, then all states of
$\tilde{\rho}_1$ in this interval must approach the equilibrium solution given
in \Eq{eq:r2}. See \Fig{f:drho1tilde_rho1} for an illustration. The same
arguments also show that $\tilde{\rho}_1=1$ [\Eq{eq:r1}] is an unstable
solution, since even a vanishingly low number density of boulders will send the
state towards $r_2$ instead. One must nevertheless still require some minimum
number of boulders in the system for the continuity description of their number
density to hold.

\section{Drag force}
\label{s:dragforce}

In this appendix we describe the implementation of drag forces in our
simulations. We let gas exert drag on the boulders following an Epstein drag
law that is linear in the velocity difference between particles and gas. The
gas velocity at the position of a particle is interpolated from the 27 nearest
grid points using spline interpolation \citep[see][]{YoudinJohansen2007}. The
Epstein drag law, with the friction time $\tau_{\rm f}$ given by
\begin{equation}
  \varOmega_{\rm K} \tau_{\rm f}^{\rm (Ep)} =
      \frac{a_2}{H} \frac{\rho_\bullet}{ \rho_{\rm g}} \, ,
\end{equation}
is valid as long as the particle radius $a_2<(9/4) \lambda$, where $\lambda$ is
the mean free path of the gas molecules. At $a_2>(9/4) \lambda$ one needs to
use instead the Stokes friction time, given by $\varOmega_{\rm K} \tau_{\rm
f}^{\rm (St)}=\varOmega_{\rm K} \tau_{\rm f}^{\rm (Ep)}\times(4/9) (a_2
/\lambda)$. The mean free path of the gas molecules can be calculated from
\begin{equation}
  \lambda = \frac{\mu}{\rho_{\rm g} \sigma_{\rm mol}} =
  \frac{\sqrt{2\pi} \mu H}{\varSigma_{\rm g} \sigma_{\rm mol}} \, ,
\end{equation}
where $\mu=3.9\times10^{-24}\,{\rm g}$ is the mean molecular weight and
$\sigma_{\rm mol}=2.0\times10^{-15}\,{\rm cm^2}$ is the cross section of
molecular hydrogen \citep{ChapmanCowling1970,Nakagawa+etal1986}. In units of
the gas scale height $H$ the mean free path can be expressed as
\begin{equation}
  \frac{\lambda}{H} = \frac{4.9 \times 10^{-9}\,{\rm
  g\,cm^{-2}}}{\varSigma_{\rm g}} \, .
\end{equation}
Assuming a minimum mass solar nebula model at $r=5\,{\rm AU}$ we have
$\varSigma_{\rm g}=150\,{\rm g\,cm^{-2}}$ and therefore a mean free path of
$\lambda/H=3.3\times10^{-11}$. The transition from Epstein to Stokes regime
thus occurs at $a_2/H=7.3\times10^{-11}$. We start our boulders with radius
$a_2/H=10^{-11}$ (with $\varOmega_{\rm K} \tau_{\rm f}=1$), so the Epstein
regime is valid up to $\varOmega_{\rm K}\tau_{\rm f}\approx7.3$, which is
already at the other side of the meter barrier. Since the focus of this paper
is the crossing of the meter barrier, we shall for simplicity model drag force
in the Epstein regime throughout and ignore the transition to the Stokes
regime. The coagulation-fragmentation equilibrium anyway has no dependence on
the assumed amount of gas in the disc -- the ratio of dust fragments to
boulders depends only on the collision speeds [\Eq{eq:rhorat_equi}], and while
the radius growth [\Eq{eq:da2dt_equi}] does scale with $\rho_{1+2}$, and thus
with $\rho_{\rm g}$ if the solids-to-gas ratio is unchanged, the evolution of
the Stokes number, ${\rm St}\propto a_2/\rho_{\rm g}$, is independent of the
gas density.

Another complication with modelling the Stokes regime is that the gas flow in
the vicinity of a boulder would lead small dust grains around the boulder
rather than onto its surface \citep{SekiyaTakeda2003}. Dust grains may however
still be able to penetrate to the boulder in case the boulder is porous and has
gas flow through it \citep{Wurm+etal2004}. We shall nevertheless limit
ourselves to the Epstein regime in this paper and leave the treatment of the
interaction of grains and boulders in the Stokes regime to a future publication.

\end{appendix}

\end{document}